\documentclass{preprint}

\usepackage{lipsum} 
\usepackage{xcolor}
\usepackage{booktabs}
\usepackage{graphicx} 
\usepackage[super,numbers,comma,sort&compress]{natbib}
\usepackage[colorlinks=true]{hyperref}
\usepackage{pdfpages}
\usepackage{tabularx}
\usepackage{array}
\usepackage{colortbl}
\usepackage{longtable}
\usepackage{xltabular}
\usepackage{ulem}
\usepackage{paralist}
\usepackage{nopageno}

\title{Environment Scan of Generative AI Infrastructure for Clinical and Translational Science}

\author[1]{Betina Idnay}
\author[2]{Zihan Xu}
\author[3]{William G. Adams}
\author[4]{Mohammad Adibuzzaman}
\author[5]{Nicholas R. Anderson}
\author[6]{Neil Bahroos}
\author[7]{Douglas S. Bell}
\author[8]{Cody Bumgardner}
\author[2,51]{Thomas Campion}
\author[9]{Mario Castro}
\author[10]{James J. Cimino}
\author[11]{I. Glenn Cohen}
\author[4]{David Dorr}
\author[12]{Peter L Elkin}
\author[13]{Jungwei W. Fan}
\author[14]{Todd Ferris}
\author[15]{David J. Foran}
\author[16]{David Hanauer}
\author[17]{Mike Hogarth}
\author[18]{Kun Huang}
\author[19]{Jayashree Kalpathy-Cramer}
\author[20]{Manoj Kandpal}
\author[21]{Niranjan S. Karnik}
\author[22]{Avnish Katoch}
\author[23]{Albert M. Lai}
\author[24]{Christophe G. Lambert}
\author[25]{Lang Li}
\author[26]{Christopher Lindsell}
\author[27]{Jinze Liu}
\author[28]{Zhiyong Lu}
\author[29]{Yuan Luo}
\author[30]{Peter McGarvey}
\author[31]{Eneida A. Mendonca}
\author[32]{Parsa Mirhaji}
\author[33]{Shawn Murphy}
\author[34]{John D. Osborne}
\author[35]{Ioannis C. Paschalidis}
\author[36]{Paul A. Harris}
\author[37]{Fred Prior}
\author[38]{Nicholas J. Shaheen}
\author[30]{Nawar Shara}
\author[39]{Ida Sim}
\author[40]{Umberto Tachinardi}
\author[41]{Lemuel R. Waitman}
\author[42]{Rosalind J. Wright}
\author[43]{Adrian H. Zai}
\author[44]{Kai Zheng}
\author[45]{Sandra Soo-Jin Lee}
\author[36]{Bradley A. Malin}
\author[1]{Karthik Natarajan}
\author[46]{W. Nicholson Price II}
\author[47]{Rui Zhang}
\author[2]{Yiye Zhang}
\author[48,*]{Hua Xu}
\author[49,*]{Jiang Bian}
\author[1,50,*]{Chunhua Weng}
\author[2,51,*]{Yifan Peng}

\affil[*]{Corresponding: Yifan Peng \url{yip4002@med.cornell.edu}, Chunhua Weng \url{cw2384@cumc.columbia.edu}, Jiang Bian \url{bianjiang@ufl.edu}, Hua Xu \url{hua.xu@yale.edu}}

\begin{document}

\maketitle

\begin{compactenum}
\footnotesize
\def\labelenumi{\textsuperscript{\arabic{enumi}}}
\item
  Department of Biomedical Informatics, Columbia University Irving Medical Center, New York, NY, USA.
\item
  Department of Population Health Sciences, Weill Cornell Medicine, New York, NY, USA.
\item
  Department of Pediatrics, Boston Medical Center, Boston, MA, USA; Chobanian \& Avedisian School of Medicine, Boston University, Boston, MA, USA.
\item
  Oregon Clinical and Translational Research Institute, Oregon Health and Science University, Portland, OR, USA.
\item
  Department of Public Health Sciences, University of California, Davis, Davis, CA, USA.
\item
  Keck School of Medicine, University of Southern California, Los Angeles, CA, USA.
\item
  Department of Medicine, David Geffen School of Medicine, University of California, Los Angeles, Los Angeles, CA, USA.
\item
  Department of Pathology and Laboratory Medicine, University of Kentucky College of Medicine, Lexington, KY, USA.
\item
  Division of Pulmonary, Critical Care and Sleep Medicine, University of Kansas School of Medicine, Kansas City, KS, USA.
\item
  Department of Biomedical Informatics and Data Science, Heersink School of Medicine, University of Alabama, Birmingham, AL, USA.
\item
  Harvard Law School, Petrie-Flom Center for Health Law Policy, Biotechnology, and Bioethics, Harvard University, Cambridge, MA, USA.
\item
  Department of Biomedical Informatics, University at Buffalo, Buffalo, NY, USA.
\item
  Center for Clinical and Translational Science, Mayo Clinic, Rochester, MN, USA.
\item
  Technology and Digital Solutions, Stanford Medicine, Stanford University, Stanford, CA, USA.
\item
  Center for Biomedical Informatics, Rutgers Cancer Institute, New Brunswick, NJ, USA.
\item
  Department of Learning Health Sciences, University of Michigan Medical School, Ann Arbor, MI, USA.
\item
  Altman Clinical and Translational Research Institute (ACTRI), University of California San Diego, La Jolla, CA, United States.
\item
  Department of Biostatistics and Health Data Science, School of Medicine, Indiana University, Indianapolis, IN, USA.
\item
  Division of Artificial Medical Intelligence in Ophthalmology, University of Colorado, Aurora, CO, USA.
\item
  Center for Clinical and Translational Science, Rockefeller University Hospital, Rockefeller University, New York, NY, USA.
\item
  AI.Health4All Center, Center for Clinical \& Translational Science, and Department of Psychiatry, University of Illinois Chicago, Chicago, IL, USA.
\item
  Department of Public Health Sciences, Penn State College of Medicine, Hershey, PA, USA.
\item
  Department of Medicine, Washington University School of Medicine, St. Louis, MO, USA.
\item
  Division of Translational Informatics, Department of Internal Medicine, University of New Mexico Health Sciences Center, Albuquerque, NM, USA.
\item
  Department of Biomedical Informatics, The Ohio State University, Columbus, OH, USA.
\item
  Duke Clinical Research Institute, Duke University, Durham, NC, USA.
\item
  Department of Population Health, Virginia Commonwealth University, Richmond, VA, USA.
\item
  Division of Intramural Research, National Library of Medicine, National Institutes of Health, Bethesda, MD, USA
\item
  Department of Preventive Medicine, Feinberg School of Medicine, Northwestern University, Chicago, IL, USA.
\item
  Georgetown-Howard Universities Center for Clinical and Translational Science, Washington, DC, USA.
\item
  Division of Biomedical Informatics, Cincinnati Children's Hospital Medical Center, Cincinnati, OH, USA.
\item
  Institute for Clinical Translational Research, Albert Einstein College of Medicine, New York, NY, USA.
\item
  Department of Neurology, Mass General Brigham, Somerville, MA, USA.
\item
  Department of Medicine, University of Alabama, Birmingham, AL, USA.
\item
  College of Engineering and Faculty of Computing \& Data Sciences, Boston University, Boston, MA, USA.
\item
  Department of Biomedical Informatics, Vanderbilt University Medical Center, Nashville, TN, USA.
\item
  Department of Biomedical Informatics, University of Arkansas for Medical Sciences, Little Rock, AR, USA.
\item
  Division of Gastroenterology and Hepatology, University of North Carolina School of Medicine, Chapel Hill, North Carolina, USA.
\item
  Department of Medicine, University of California, San Francisco, San Francisco, CA, USA.
\item
  Department of Biostatistics, Health Informatics and Data Sciences, University of Cincinnati College of Medicine, Cincinnati, OH, USA.
\item
  Department of Biomedical Informatics, Biostatistics, and Medical Epidemiology, School of Medicine, University of Missouri, Columbia, MO, USA.
\item
  Department of Public Health, Icahn School of Medicine at Mount Sinai, New York, NY, USA.
\item
  Division of Health Informatics and Implementation Science, Department of Population and Quantitative Health Sciences, UMass Chan Medical School, Worcester, MA, USA.
\item
  Department of Informatics, University of California, Irvine, Irvine, CA, USA.
\item
  Department of Medical Humanities and Ethics, Columbia University, New York, NY, USA.
\item
  Michigan Law School, University of Michigan, Ann Arbor, MI, USA.
\item
  Division of Computational Health Sciences, Medical School, University of Minnesota, Minneapolis, MN, USA.
\item
  Department of Biomedical Informatics and Data Science, Yale School of Medicine, Yale University, New Haven, CT, USA.
\item
  Department of Neurology and McKnight Brain Institute, College of Medicine, University of Florida, Gainesville, FL, USA.
\item
  The Irving Institute for Clinical and Translational Research, Columbia University, New York, NY, USA.
\item
  Clinical and Translational Science Center, Weill Cornell Medicine, New York, NY, USA
\end{compactenum}

\newpage

\begin{abstract}
This study reports a comprehensive environmental scan of the generative AI (GenAI) infrastructure in the national network for clinical and translational science across 36 institutions supported by the Clinical and Translational Science Award (CTSA) Program led by the National Center for Advancing Translational Sciences (NCATS) of the National Institutes of Health (NIH) at the United States. With the rapid advancement of GenAI technologies, including large language models (LLMs), healthcare institutions face unprecedented opportunities and challenges. This research explores the current status of GenAI integration, focusing on stakeholder roles, governance structures, and ethical considerations by administering a survey among leaders of health institutions (i.e., representing academic medical centers and health systems) to assess the institutional readiness and approach towards GenAI adoption. Key findings indicate a diverse range of institutional strategies, with most organizations in the experimental phase of GenAI deployment. The study highlights significant variations in governance models, with a strong preference for centralized decision-making but notable gaps in workforce training and ethical oversight. Moreover, the results underscore the need for a more coordinated approach to GenAI governance, emphasizing collaboration among senior leaders, clinicians, information technology staff, and researchers. Our analysis also reveals concerns regarding GenAI bias, data security, and stakeholder trust, which must be addressed to ensure the ethical and effective implementation of GenAI technologies. This study offers valuable insights into the challenges and opportunities of GenAI integration in healthcare, providing a roadmap for institutions aiming to leverage GenAI for improved quality of care and operational efficiency.
\end{abstract}


\section{INTRODUCTION}

The burgeoning advancement of generative AI (GenAI) provides transformative potential for healthcare systems globally. GenAI employs computational models to generate new content based on patterns learned from existing data. These models, exemplified by large language models (LLMs), can produce content across various modalities such as text, images, video, and audio.\cite{Huang2023-wd, Matsubayashi2024-xt, Saaran2021-dp, Kazerouni2023-tn, Wei2021-iz}
Its ability to generate human comprehensible text enabled the exploration of diverse applications in healthcare that involve the sharing and dissemination of expert knowledge, ranging from clinical decision support to patient engagement.\cite{Brown2020-xz, Touvron2023-hz}
Integrating GenAI into healthcare can enhance diagnostic accuracy, personalized treatment plans, and operational efficiencies. For instance, GenAI-driven diagnostic tools can analyze medical images and electronic health records (EHRs) to detect diseases, often surpassing the accuracy of human experts.\cite{Alowais2023-xy, Topol2019-gl, Kim2024-yq, Rajpurkar2022-fi, Hao2024-ow, Amini2024-fp}
GenAI applications can streamline administrative processes, reduce clinicians' documentation burden, and enable them to spend more time on direct patient care.\cite{Jiang2017-bd, Secinaro2021-mz}
However, implementing GenAI technologies in healthcare has several challenges. Issues such as trustworthiness, data privacy, algorithmic bias, and the need for robust regulatory frameworks are critical considerations that must be addressed to ensure the responsible and effective use of GenAI.\cite{Obermeyer2019-ia, Polevikov2023-bf}

Given these promising advancements and associated challenges, understanding the current institutional infrastructure for implementing GenAI in healthcare is crucial. Various stakeholders (e.g., clinicians, patients, researchers, regulators, industry professionals) have different roles and responsibilities in GenAI implementation, ranging from ensuring patient safety and data security to driving innovation and regulatory compliance, and may hold varying attitudes toward GenAI applications that influence their acceptance and utilization of these technologies. Failure to consider these diverse perspectives may hinder the widespread adoption and effectiveness of GenAI technologies.

Previous studies have examined stakeholder perspectives on AI adoption to some extent. For example, Scott et al. found that while various stakeholders generally had positive attitudes towards AI in healthcare, especially those with direct experience, significant concerns persisted regarding privacy breaches, personal liability, clinician oversight, and the trustworthiness of AI-generated advice.\cite{Scott2021-iu}
These concerns are reflective of AI technologies in general. Specific to GenAI, Spotnitz et al. surveyed healthcare providers and found that while clinicians were generally positive about using LLMs for assistive roles in clinical tasks, they had concerns about generating false information and propagating training data bias.\cite{Spotnitz2024-rm}

Despite these insights, there remains a gap in understanding the infrastructure required for GenAI integration in healthcare institutions, particularly from the perspective of institutional leadership. The Clinical and Translational Science Awards (CTSA) Program, funded by the National Center for Advancing Translational Sciences (NCATS) of the National Institutes of Health (NIH) at the United States, supports a nationwide consortium of medical research institutions at the forefront of clinical and translational research and practice.\cite{Liverman2013-de}
By examining the GenAI infrastructure within CTSA institutions, we can gain valuable insights into how GenAI is being adopted into cutting-edge research environments and help set benchmarks for the broader healthcare community. Furthermore, understanding the challenges faced by CTSA institutions in this context is crucial for developing strategies that promote fair and accessible GenAI implementation.\cite{Alowais2023-xy, Yin2021-so}

In this study, we aim to conduct an environmental scan of the infrastructure for GenAI within CTSA institutions by surveying CTSA leaders to comprehensively understand its current integration status. We also highlight opportunities and challenges in achieving equitable GenAI implementation in healthcare by identifying key stakeholders, governance structures, and ethical considerations. We acknowledge the dual roles that respondents may represent, whether in their capacity as leaders within academic institutions (i.e., CTSA), healthcare systems, or both. Hence, we use the term ``healthcare institutions'' to encompass the broad range of leadership representation and capture a more complete picture of GenAI integration across research-focused and healthcare-delivery institutions. The insights gained from this study can inform the development of national policies and guidelines to ensure the ethical use of GenAI in healthcare; identifying successful GenAI implementation strategies can serve as best practices for other institutions; highlighting gaps in the current GenAI infrastructure can guide future investments and research priorities; and ultimately, a robust GenAI infrastructure can enhance patient care through more accurate diagnoses, personalized treatments, and efficient healthcare delivery.

\section{RESULTS}

The US CTSA network contains over 60 hubs. We sent email invitations to 64 CTSA leaders, each responding on behalf of a unique CTSA site, with 42 confirming participation. Ultimately, we received 36 complete responses, yielding an 85.7\% completion rate. Only fully completed responses were included in the analysis, as the six unfinished responses had 0-65\% progress and were excluded. The survey questions are available in the \textbf{Supplementary File \ref{sec:questionnaire}}. Of the 36 completed responses, 15 (41.7\%) represented only a CTSA, and 21 (58.3\%) represented a CTSA and its affiliated hospital.

\subsection{Stakeholder Identification and Roles}\label{stakeholder-identification-and-roles}

\begin{figure}
    \centering
    \includegraphics[width=\linewidth]{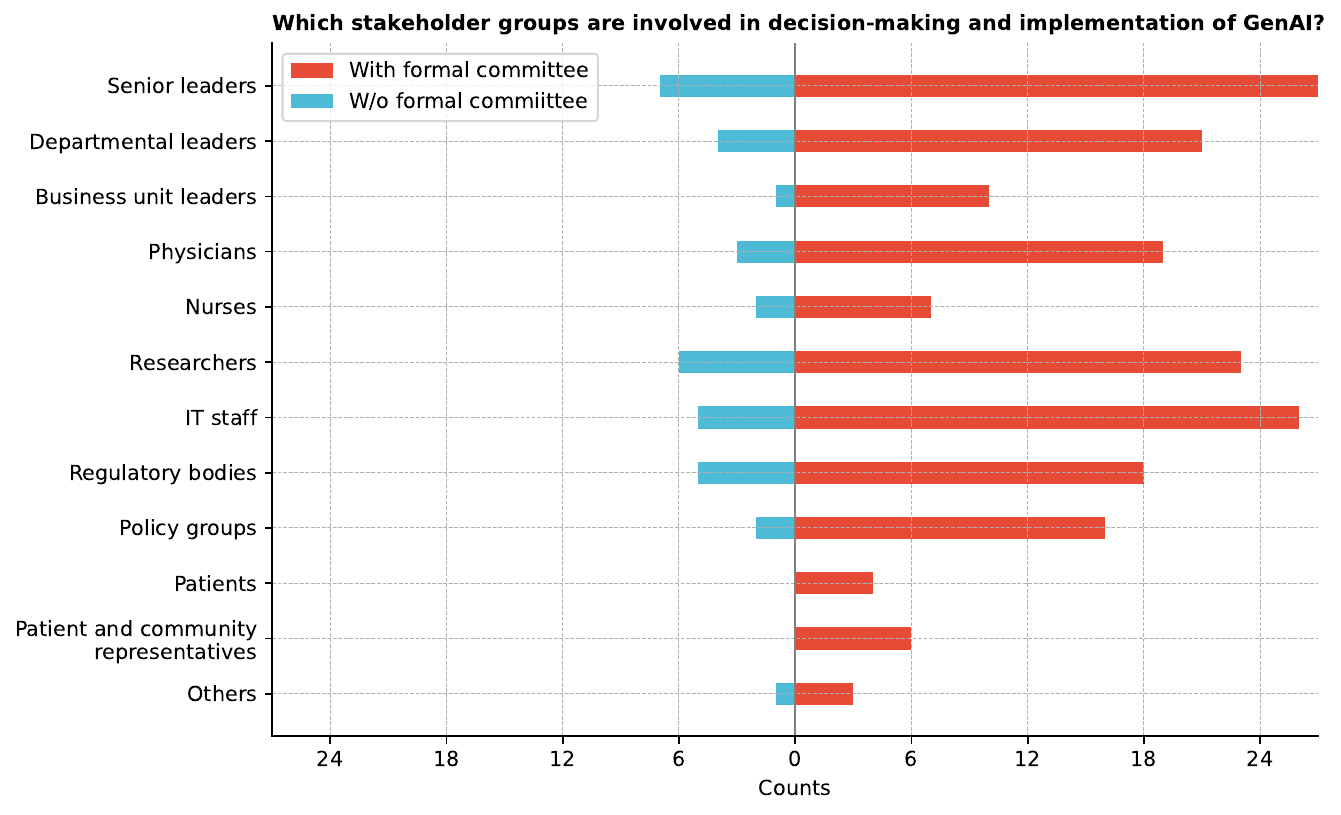}
    \caption{Which stakeholder groups are involved in your organization's decision-making and implementation of GenAI?}
    \label{fig:q2}
\end{figure}

\textbf{Figure \ref{fig:q2}} shows that senior leaders were the most involved in GenAI decision-making (94.4\%), followed by information technology (IT) staff, researchers, and physicians. Cochran's Q test revealed significant differences in stakeholder involvement (Q = 165.9, p \textless{} 0.0001). Post-hoc McNemar tests (see Methods) with Bonferroni correction showed senior and departmental leaders were significantly more involved than business unit leaders, nurses, patients, and community representatives (all corrected p \textless{} 0.0001). Nurses were also less engaged than IT staff (corrected p \textless{} 0.0001) (See \textbf{Supplementary Table \ref{sup tab:stakeholder}}).


We further split our analysis based on whether institutions have formal committees or task forces overseeing GenAI governance to provide insights into how governance models may impact GenAI adoption. 77.8\% (28/36) respondents reported having formal committees or task forces overseeing GenAI governance, 19.4\% (7/36) did not, and 2.8\% (1/36) were unsure. We grouped those without formal committees for analysis to simplify the comparison and focus on clear distinctions between institutions with and without established governance structures. Institutions without formal committees did not involve patients and community representatives as stakeholders in the decision-making and implementation of GenAI (\textbf{Figure \ref{fig:q2})}.

\begin{figure}
    \centering
    \includegraphics[width=\linewidth]{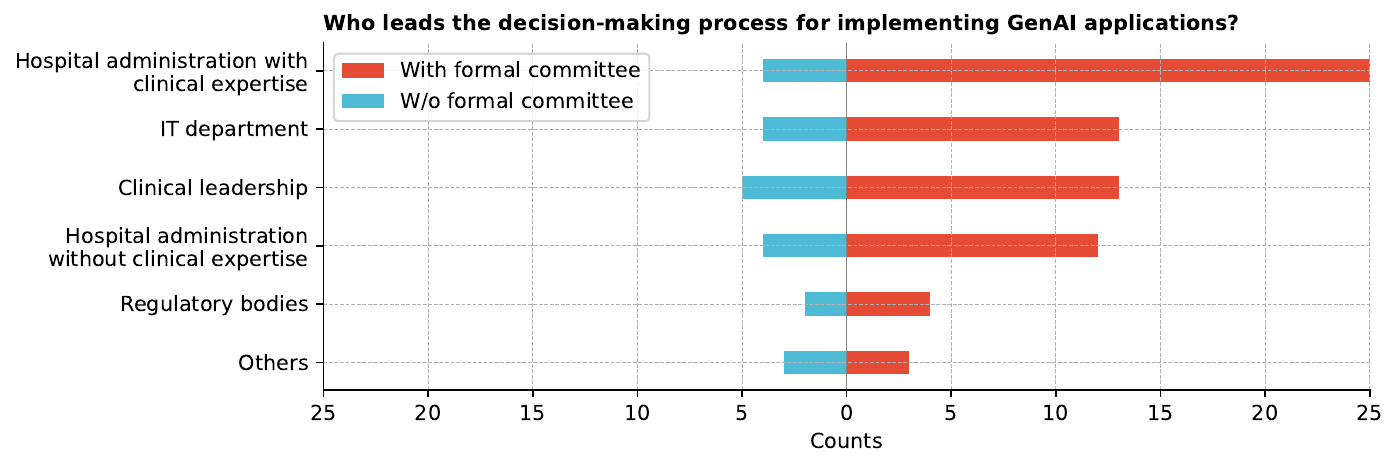}
    \caption{Who leads the decision-making process for implementing GenAI applications in your organization?}
    \label{fig:q3}
\end{figure}

Further, the decision-making process for implementing GenAI (\textbf{Figure \ref{fig:q3}}) was primarily led by cross-functional committees (80.6\%), with clinical leadership also playing a key role (50.0\%). Institutions without formal committees were led more by clinical leadership. Specific mentions include the dean, CTSA and innovation teams, researchers, and health AI governance committees. Cochran's Q test revealed significant differences in leadership involvement (Q = 46.8, p \textless{} 0.0001), especially between cross-functional committees and both regulatory bodies and other stakeholders (corrected p \textless{} 0.0001) (See \textbf{Supplementary Table~\ref{sup tab:decision}}).

\subsection{Decision-Making and Governance Structure}\label{decision-making-and-governance-structure}

\begin{figure}
    \centering
    \includegraphics[width=\linewidth]{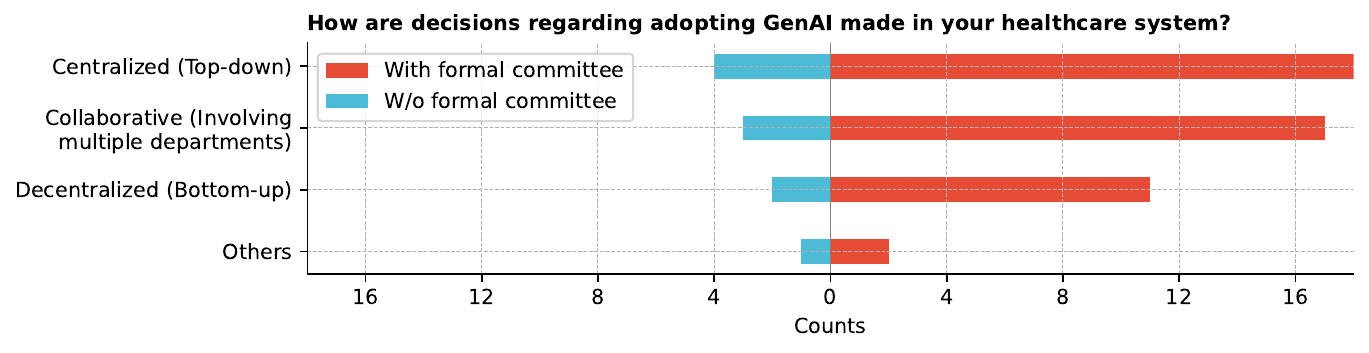}
    \caption{How are decisions regarding adopting GenAI made in your healthcare institution?}
    \label{fig:q4}
\end{figure}

The decision-making process for adopting GenAI in healthcare institutions varied (\textbf{Figure \ref{fig:q4}}). A centralized (top-down) approach was used by 61.1\% (22/36) of respondents, while 8.3\% (3/36) mentioned alternative methods, such as decisions based on the tool's nature or a mix of centralized and decentralized approaches.

\begin{figure}
    \centering
    \includegraphics[width=.8\linewidth]{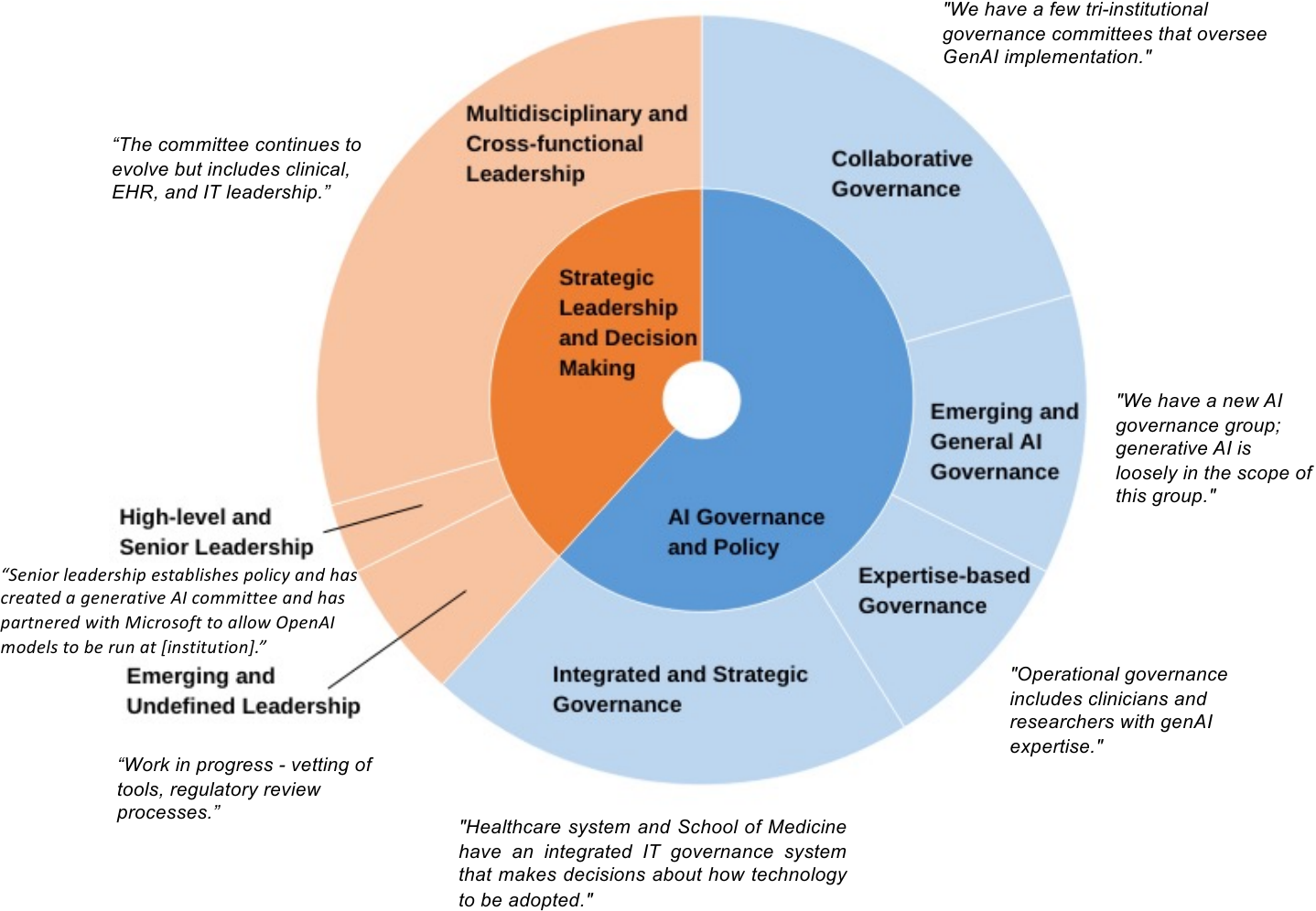}
    \caption{Thematic analysis of governance and leadership structures in GenAI deployment across CTSA institutions with featured responses.}
    \label{fig:thematic}
\end{figure}

Thematic analysis of statements about governance structures in organizations with formal committees identified two major themes (\textbf{Figure \ref{fig:thematic}}). ``AI Governance and Policy'' reflects institutions' structured approaches to ensure responsible GenAI implementation. Institutions often establish multidisciplinary committees to integrate GenAI policies with existing frameworks, aligning AI deployment with organizational goals and regulatory requirements and focusing on legal and ethical compliance. ``Strategic Leadership and Decision Making'' highlights the crucial role of leadership in GenAI initiatives. High-level leaders drive GenAI integration through strategic planning and resource allocation, with integrated teams from IT, research, and clinical care fostering a culture of innovation and collaboration. Excerpts on these governance practices are detailed in the \textbf{Supplementary Table \ref{sup tab:experts}}.

\subsection{Regulatory and Ethical Considerations}\label{regulatory-and-ethical-considerations}

\begin{figure}
    \centering
    \includegraphics[width=\linewidth]{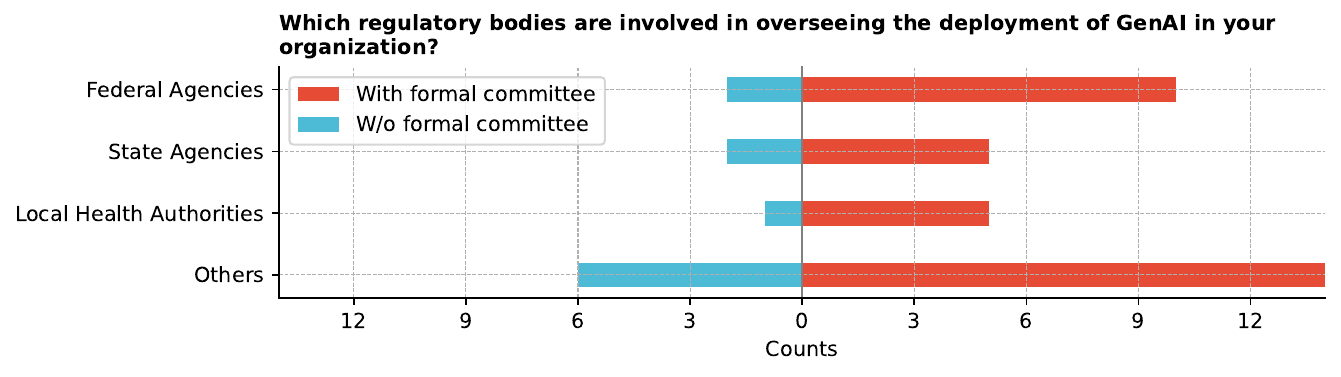}
    \caption{Which regulatory bodies are involved in overseeing the deployment of GenAI in your organization?}
    \label{fig:q6}
\end{figure}

Regulatory body involvement in GenAI deployment varied widely across institutions (\textbf{Figure \ref{fig:q6}}). Federal agencies were engaged in 33.3\% (12/36) of organizations. A significant portion (55.6\%) identified other bodies, including institutional review boards (IRBs), ethics committees, community advocates, and state agencies. Internal governance committees and university task forces were also explicitly mentioned.

\begin{figure}
    \centering
    \includegraphics[width=\linewidth]{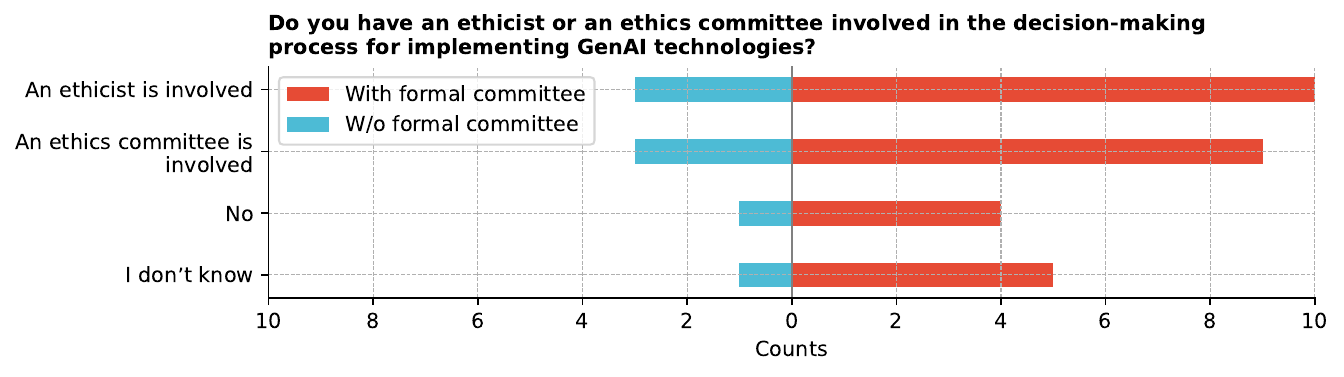}
    \caption{Do you have an ethicist or an ethics committee involved in the decision-making process for implementing GenAI technologies in your organization?}
    \label{fig:q7}
\end{figure}

\begin{figure}
    \centering
    \includegraphics[width=\linewidth]{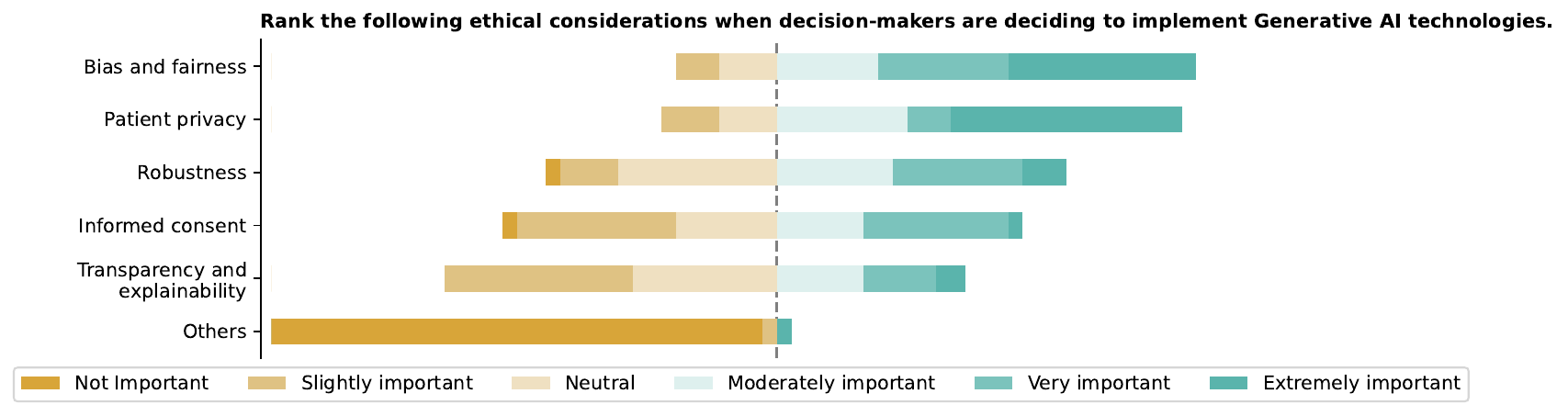}
    \caption{Please rank the following ethical considerations from most important (1) to least important (6) when decision-makers are deciding to implement GenAI technologies.}
    \label{fig:q8}
\end{figure}

Regarding ethical oversight (\textbf{Figure \ref{fig:q7}}), 36.1\% (13/36) of respondents reported an ethicist's involvement in GenAI decision-making; 27.8\% (10/36) mentioned an ethics committee, while 19.4\% (7/36) reported neither, and 16.7\% (6/36) were unsure. Ethical considerations were ranked based on importance (\textbf{Figure \ref{fig:q8}}), with ``Bias and fairness'' (mean rank 2.31) and ``Patient Privacy'' (mean rank 2.36) being the top priorities.

\subsection{Stage of Adoption}\label{stage-of-adoption}

\begin{figure}
    \centering
    \includegraphics[width=\linewidth]{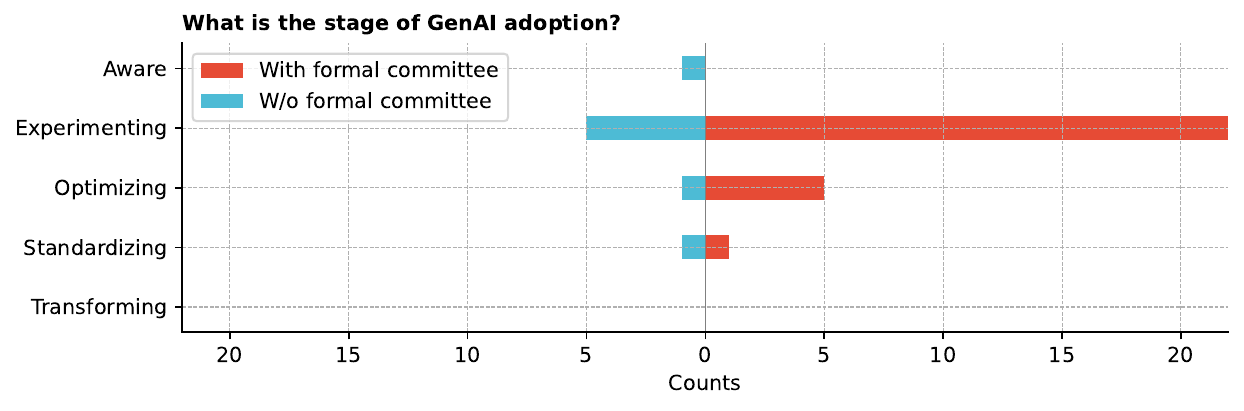}
    \caption{What is the stage of GenAI adoption in your organization?}
    \label{fig:q9}
\end{figure}

\begin{figure}
    \centering
    \includegraphics[width=\linewidth]{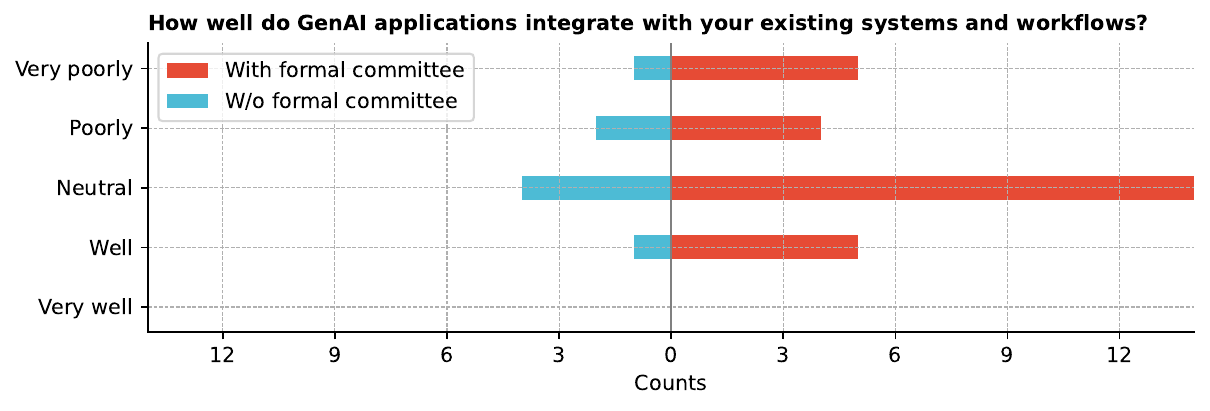}
    \caption{How well do GenAI applications integrate with your existing systems and workflows?}
    \label{fig:q10}
\end{figure}

Institutions were at varying stages of GenAI adoption (\textbf{Figure \ref{fig:q9}}), with 75.0\% (27/36) in the experimentation phase, focusing on exploring AI's potential, building skills, and identifying areas for value addition. Integrating existing systems and workflows was met with mixed responses (\textbf{Figure \ref{fig:q10}}), with 50.0\% (18/36) rating it as neutral.

\begin{figure}
    \centering
    \includegraphics[width=\linewidth]{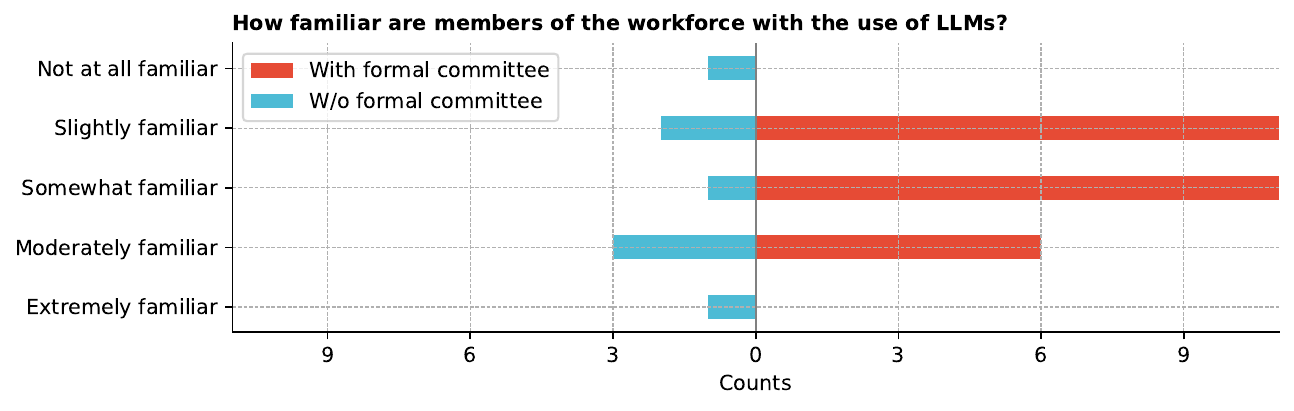}
    \caption{How familiar are members of the workforce with the use of LLMs in your organization?}
    \label{fig:q11}
\end{figure}

\begin{figure}
    \centering
    \includegraphics[width=\linewidth]{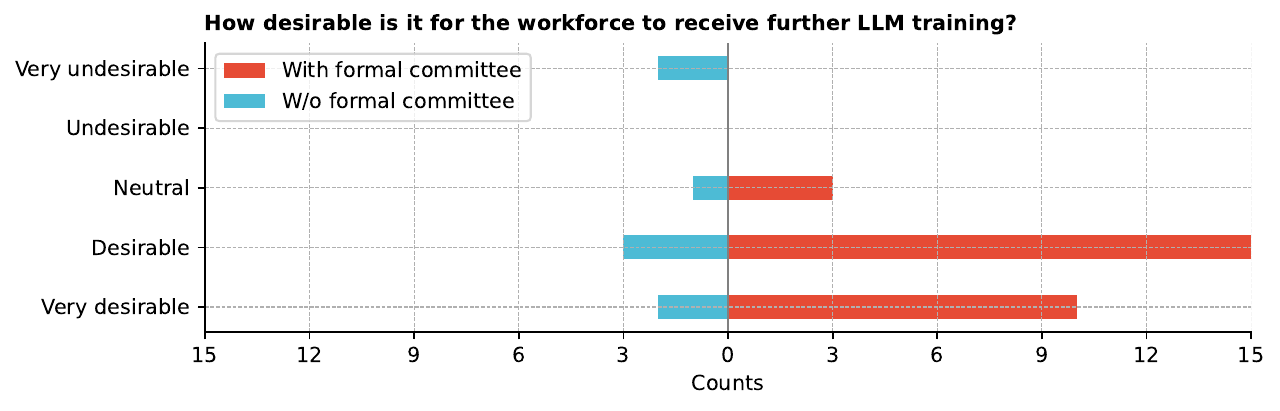}
    \caption{How desirable is it for the workforce to receive further LLM training?}
    \label{fig:q13}
\end{figure}

Workforce familiarity with large LLMs also varied (\textbf{Figure \ref{fig:q11}}), with 36.1\% (13/36) of respondents reporting slight familiarity and 25.0\% (9/36) reporting moderate familiarity. Workforce training on LLMs was uneven, with only 36.1\% (13/36) having received training, while 44.4\% (16/36) considered but did not receive training, and 19.4\% (7/36) neither received nor considered training. The demand for further training was evident, with 83.3\% (30/36) finding it desirable or even more (\textbf{Figure \ref{fig:q13}}). The respondents who indicated receiving further LLM training for their workforce was undesirable were from institutions without a formal committee.

Vendor collaboration was crucial, with 69.4\% (25/36) of institutions partnering with multiple vendors, ranging from one to twelve, to implement GenAI solutions. Notable vendors included Dax Co-pilot, Microsoft Azure AI, Amazon Web Services, Epic Systems, and various startups. Some respondents noted that discussions are often confidential or lack comprehensive information on enterprise-wide vendor engagements. Additionally, 25.0\% (9/36) have considered vendor collaboration but have not engaged, while only 5.6\% (2/36) have neither considered nor pursued such partnerships.

\subsection{Budget Trends}\label{budget-trends}

\begin{figure}
    \centering
    \includegraphics[width=\linewidth]{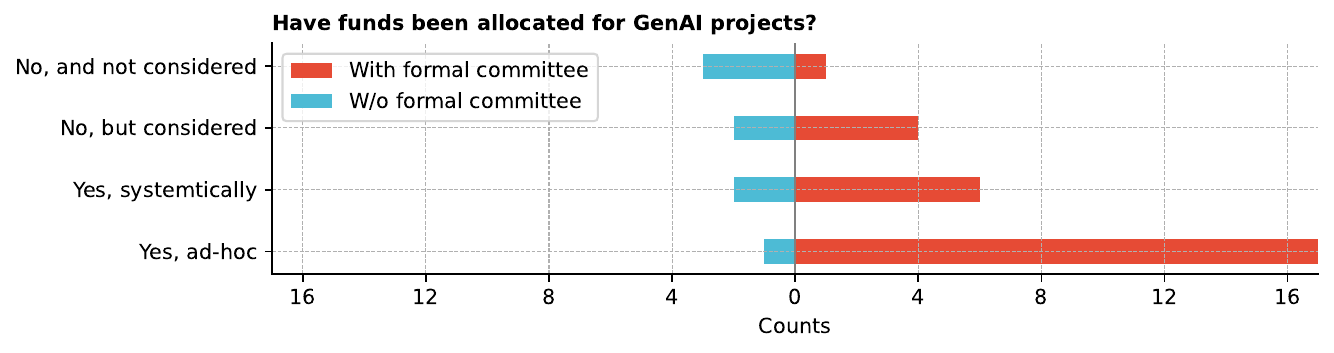}
    \caption{Have funds been allocated for GenAI projects?}
    \label{fig:q15}
\end{figure}

\begin{figure}
    \centering
    \includegraphics[width=\linewidth]{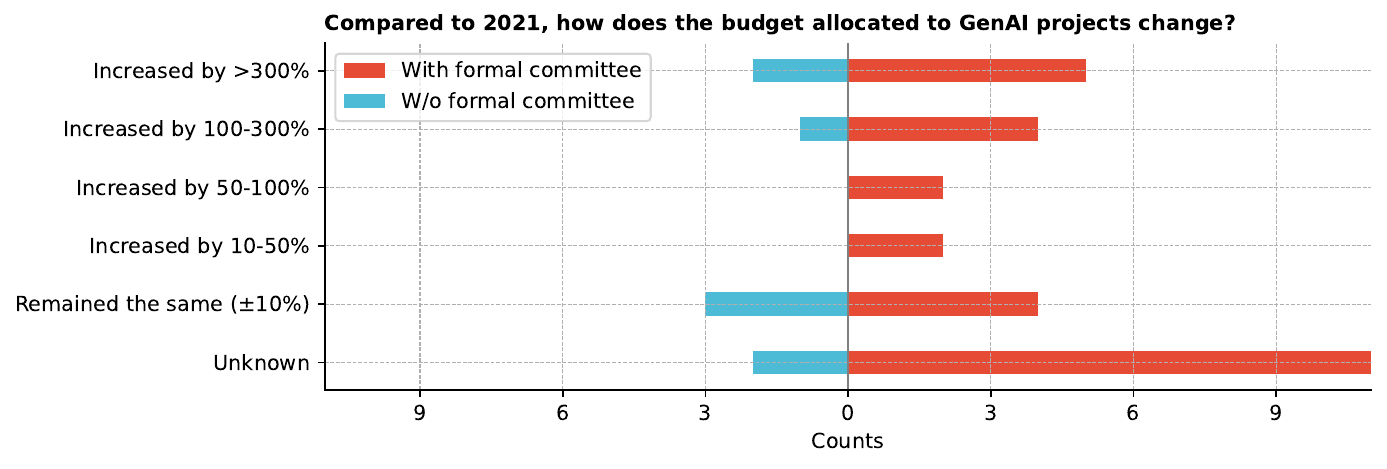}
    \caption{Compared to 2021, how does the budget allocated to GenAI projects in your organization change? }
    \label{fig:q16}
\end{figure}

Regarding funds allocation for GenAI projects, 50.0\% (18/36) of respondents reported that ad-hoc funding was allocated mostly from institutions with formal committees (\textbf{Figure \ref{fig:q15}}). Most institutions without formal committees reported that no funds had been allocated for GenAI projects (62.5\%; 5/8). Since 2021, 36.1\% (13/36) of respondents were unsure about budget changes, 19.4\% (7/36) noted the budget remained roughly the same, and 44.5\% reported budget increases ranging from 10\% to over 300\% (\textbf{Figure \ref{fig:q16}}).

\subsection{Current LLM usage}\label{current-llm-usage}

\begin{figure}
    \centering
    \includegraphics[width=\linewidth]{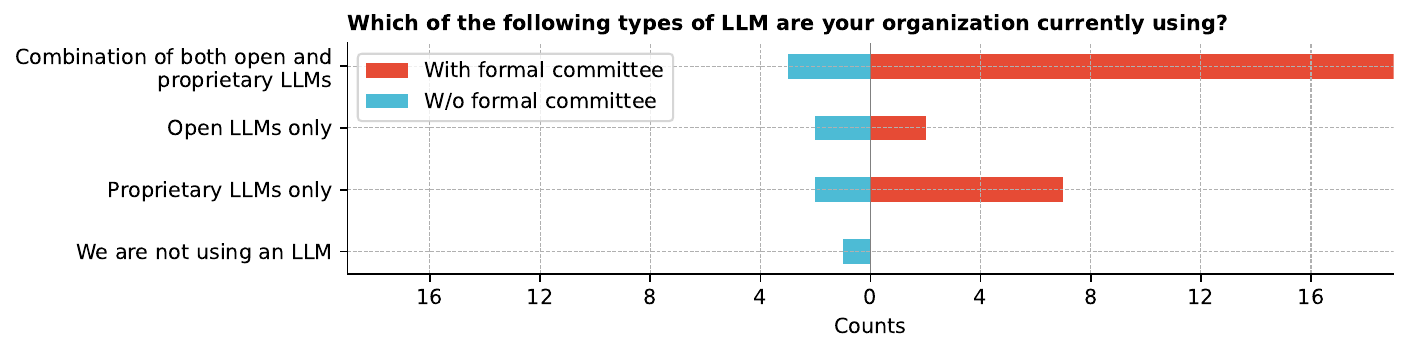}
    \caption{Which of the LLMs are you currently using?}
    \label{fig:q17}
\end{figure}

\begin{figure}
    \centering
    \includegraphics[width=\linewidth]{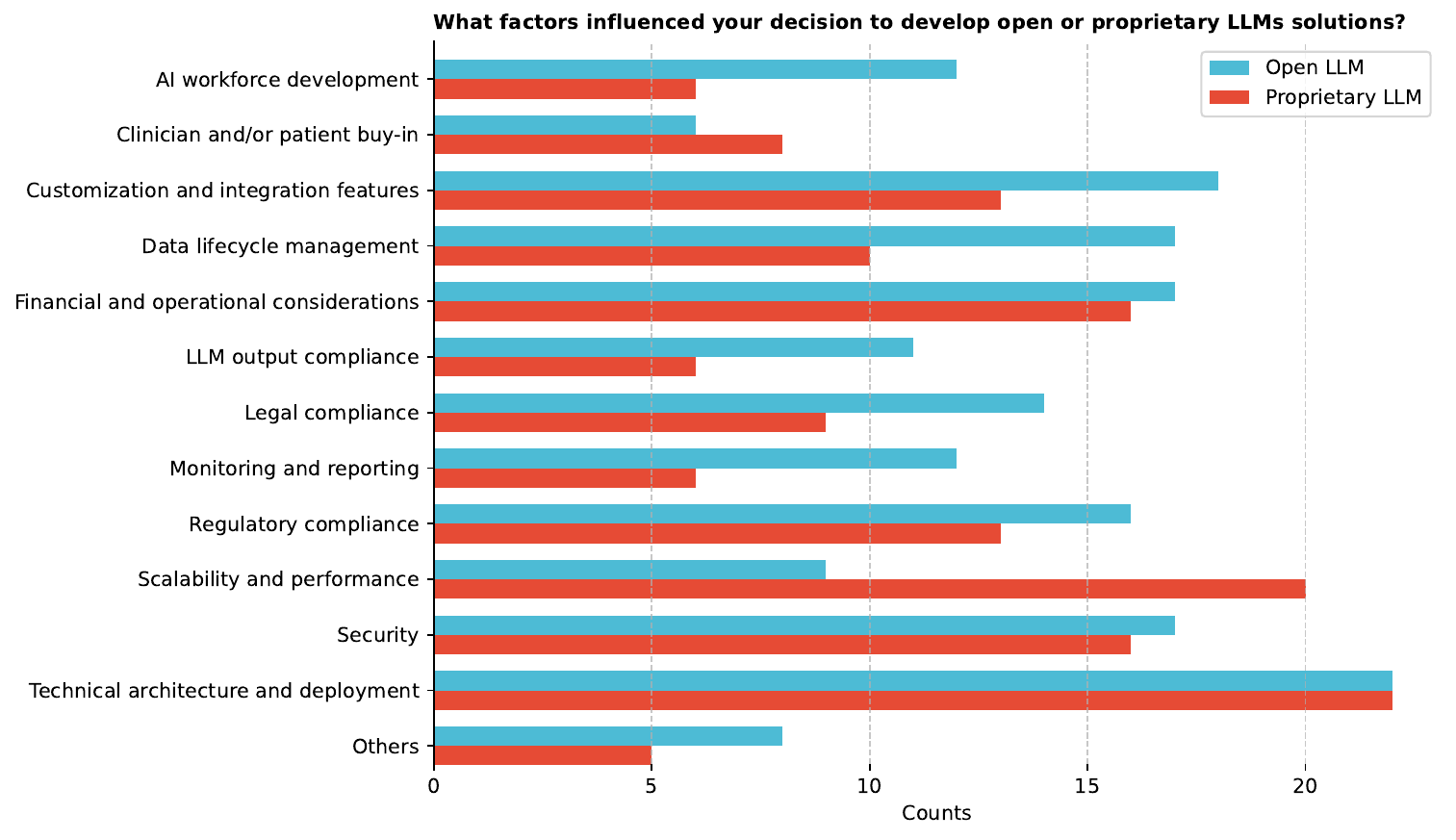}
    \caption{You indicated that your organization is using open LLMs (blue) or proprietary LLMs (red). What factors influenced your decision to develop internally/to go with commercial solutions? }
    \label{fig:q19}
\end{figure}

Institutions were adopting LLMs with varied strategies (\textbf{Figure \ref{fig:q17}}), with 61.1\% (22/36) using a combination of both open and proprietary LLMs, 11.1\% (4/36) using open LLMs only, and 25.0\% (9/36) using proprietary LLMs only. Only 2.8\% (1/36) reported not using any LLMs. Significant differences exist (Q = 28.7, p \textless{} 0.0001) between the types of LLMs used. Post-hoc tests revealed significant differences between using open and proprietary LLMs versus open LLMs only (corrected p = 0.0032) (See \textbf{Supplementary Table~\ref{sup tab:llm use}}), indicating a notable preference for combining different LLM types in some institutions. No significant differences were found among specific open or proprietary LLM types (Q = 2.4, p = 0.4936), suggesting that institutions did not exhibit strong preferences between particular open or proprietary LLM models. Institutions developing open LLMs prioritized technical architecture and deployment (61.1\%), followed by customization and integration features (50.0\%, \textbf{Figure \ref{fig:q19}}). Some institutions focused on research and experimentation, comparing open to proprietary LLMs, with interests in medical education and cost-effectiveness. Technical architecture and deployment are prioritized over clinician or patient buy-in (corrected p = 0.0024) (See \textbf{Supplementary Table \ref{sup tab:open LLM}}).

\begin{figure}
    \centering
    \includegraphics[width=\linewidth]{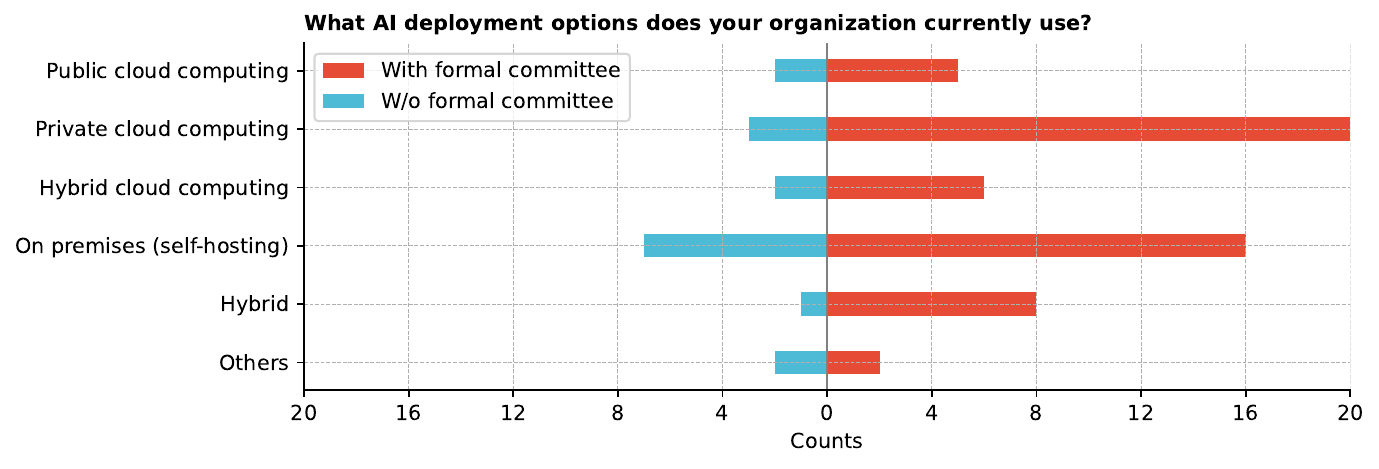}
    \caption{What AI deployment options does your organization currently use?}
    \label{fig:q18}
\end{figure}

Regarding GenAI deployment (\textbf{Figure \ref{fig:q18}}), private cloud and on-premises self-hosting were the most common approaches (both 63.9\%), suggesting that most institutions have both approaches but do not take a hybrid approach. Some institutions specified using local supercomputing resources or statewide high-performance computing infrastructure. Statistical analysis (Q = 42.6, p \textless{} 0.0001) indicated a preference for more controlled environments, with private cloud and on-premises self-hosting significantly more favored than public cloud (corrected p = 0.0022 and p = 0.0060, respectively) (See \textbf{Supplementary Table \ref{sup tab:factor commercial}}).

For institutions adopting proprietary LLMs, the critical factors for decision-making include technical architecture and deployment (61.1\%), and scalability and performance (\textbf{Figure \ref{fig:q18}}). Respondents noted the importance of ease of deployment, especially in partnerships with vendors like Epic Systems and Oracle, and the advantage of existing Health Insurance Portability and Accountability Act (HIPAA) Business Associate Agreements with providers like Microsoft. Statistical analysis (Q = 57.4, p \textless{} 0.0001) revealed significant differences, particularly between technical architecture and deployment and monitoring and reporting and AI workforce development (both corrected p = 0.0113). Scalability and performance were significantly more prioritized than LLM output compliance and AI monitoring and reporting (corrected p-values = 0.0405) (See \textbf{Supplementary Table \ref{sup tab:proprietary LLM}}).

Finally, LLMs were applied across diverse domains, with common uses in biomedical research (66.7\%), medical text summarization (66.67\%), and data abstraction (63.9\%, \textbf{Figure \ref{fig:q21}}). Co-occurrence analysis showed frequent overlaps in these areas (See \textbf{Supplementary Table \ref{sup tab:factors open}}). Medical imaging analysis was the most common use case for institutions without formal committees overseeing GenAI governance. Significant differences were observed in using LLMs for data abstraction compared to drug development, machine translation, and scheduling and between biomedical research and drug development, machine translation, and scheduling (corrected p-values \textless{} 0.05) (See \textbf{Supplementary Table \ref{sup tab:LLM use}}).
\begin{figure}
    \centering
    \includegraphics[width=\linewidth]{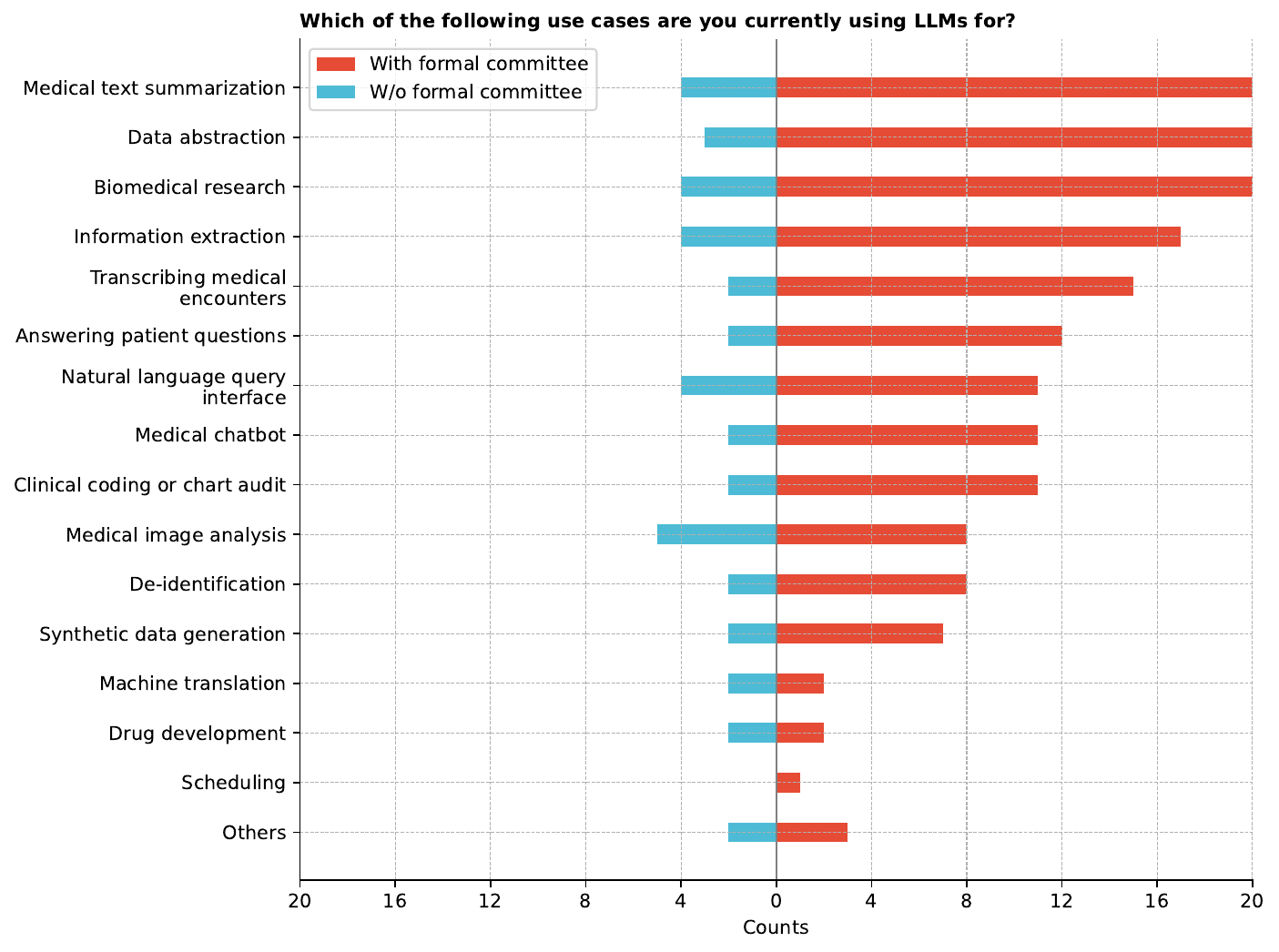}
    \caption{Which of the following use cases are you currently using LLMs for?}
    \label{fig:q21}
\end{figure}

\subsection{LLM Evaluation }\label{llm-evaluation}

Respondents prioritized accuracy and reproducible and consistent answers when evaluating LLMs for healthcare (\textbf{Figure \ref{fig:q25}}), each receiving the highest mean rating of 4.5 (See \textbf{Supplementary Table \ref{sup tab:evaluating mean-ratings}}). Healthcare-specific models and security and privacy risks were also deemed important, though responses varied. An Analysis of Variance (ANOVA) test revealed significant differences among the importance ratings (F = 3.4, p = 0.0031). Post-hoc Tukey's honestly significant difference (HSD) tests showed a significant difference between accuracy, and explainability and transparency (p = 0.0299).
\begin{figure}
    \centering
    \includegraphics[width=\linewidth]{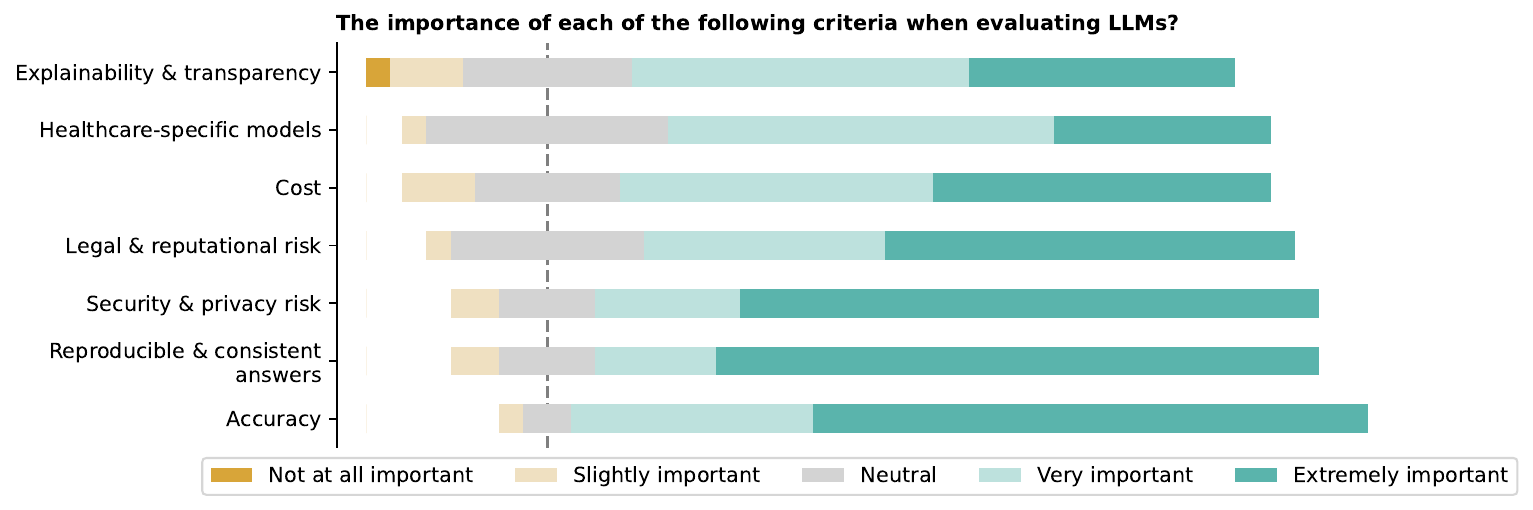}
    \caption{On a scale from 1 to 5, please rate the importance of each of the following criteria when evaluating LLMs. 1 means "Not at all Important," and 5 means "Extremely Important".}
    \label{fig:q25}
\end{figure}

Regarding potential roadblocks to adopting GenAI in healthcare, regulatory compliance issues were rated as the most significant concern, with a mean rating of 4.2 (\textbf{Figure \ref{fig:q26})} (Mean Rating See \textbf{Supplementary Table \ref{sup tab:genai mean-ratings}}). While `Too expensive' and `Not built for healthcare and life science' were less of a concern, they still posed challenges for some respondents, though there are no significant differences among these ratings (F = 2.0, p = 0.0606).
\begin{figure}
    \centering
    \includegraphics[width=\linewidth]{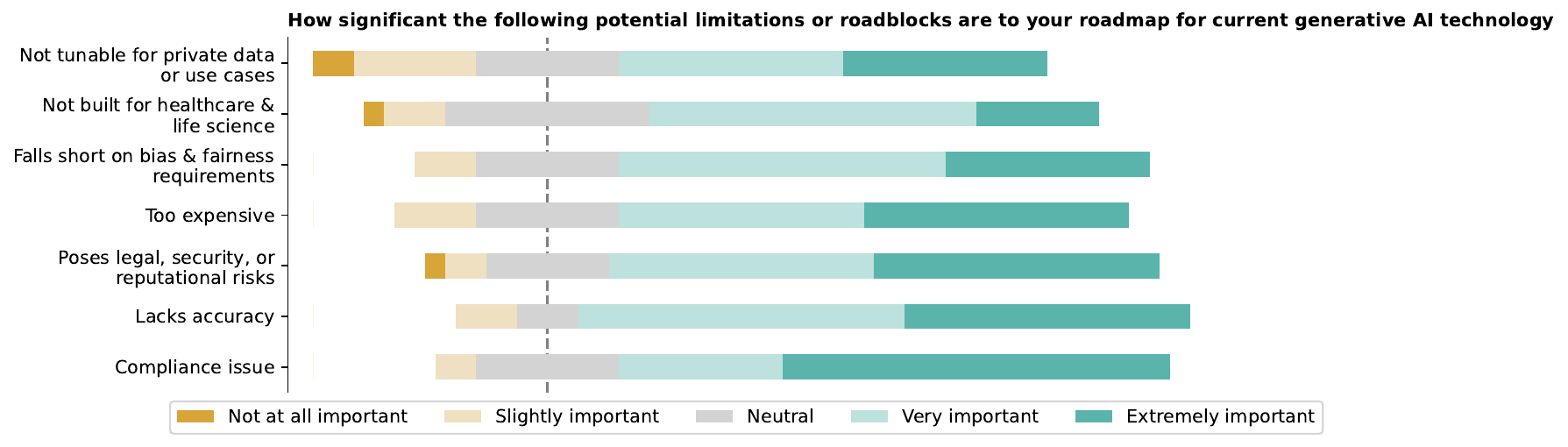}
    \caption{On a scale of 1 to 5, please rate how significant the following potential limitations or roadblocks are to your roadmap for current generative AI technology, with 1 being not important and 5 being very important. }
    \label{fig:q26}
\end{figure}

\subsection{Projected Impact}\label{projected-impact}

\begin{figure}
    \centering
    \includegraphics[width=\linewidth]{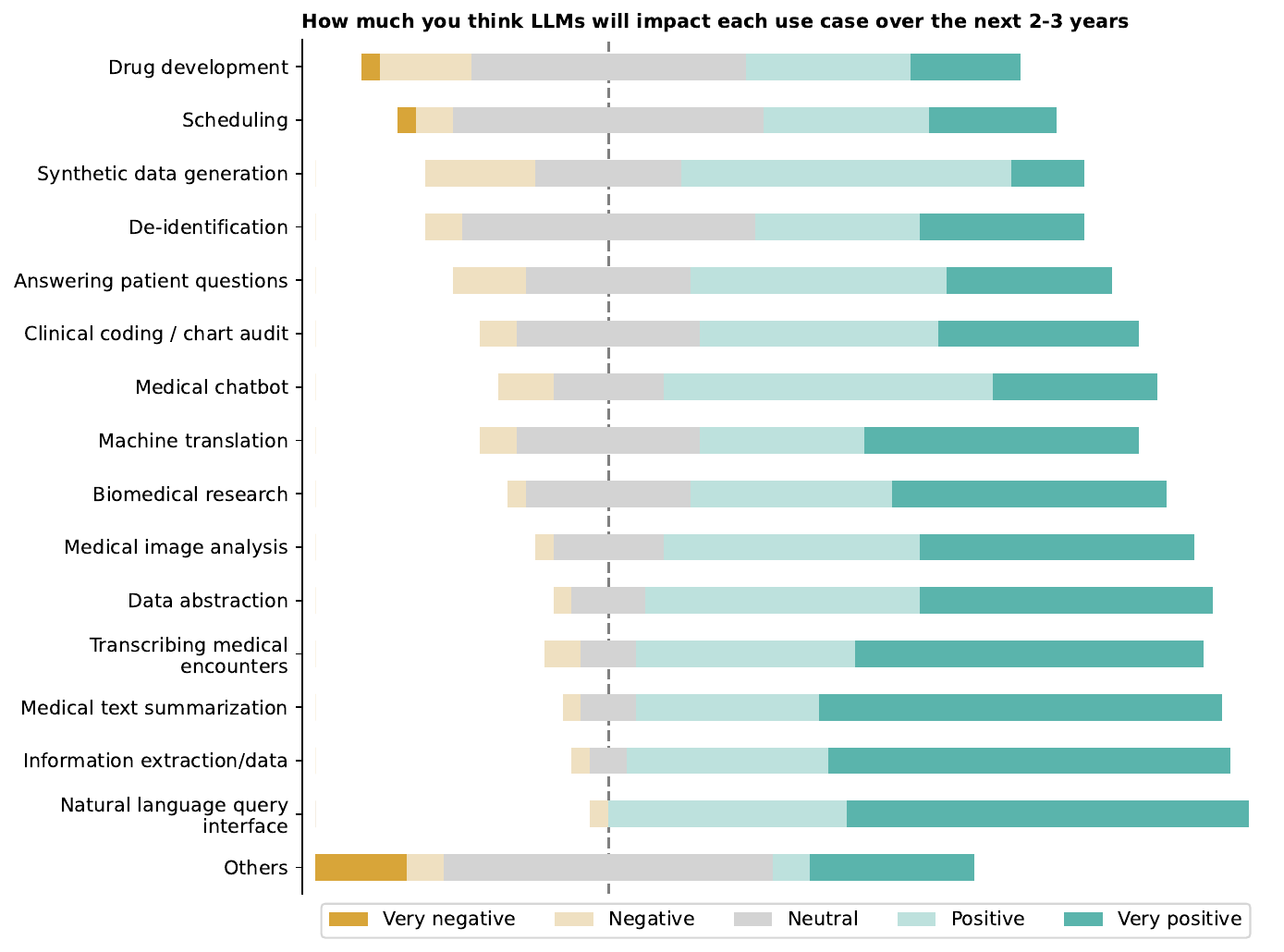}
    \caption{On a scale of 1 to 5, please rate how much you think LLMs will impact each use case over the next 2-3 years. 1 means very negative, and 5 means very positive.}
    \label{fig:q22}
\end{figure}

Participants rated the anticipated impact of LLMs on various use cases over the next 2-3 years (\textbf{Figure~\ref{fig:q22}}), with the highest mean ratings for natural language query interface, information extraction, and medical text summarization (4.5 each), followed by transcribing medical encounters (4.3). Data abstraction (4.3) and medical image analysis (4.2) were also highly rated, while synthetic data generation, scheduling (3.5 each), and drug development (3.4) received lower ratings (See \textbf{Supplementary Table \ref{sup tab:impact mean-ratings}}). Additional use cases, such as medical education and decentralized clinical trials, suggest an expanding scope for LLM applications.

\begin{figure}
    \centering
    \includegraphics[width=\linewidth]{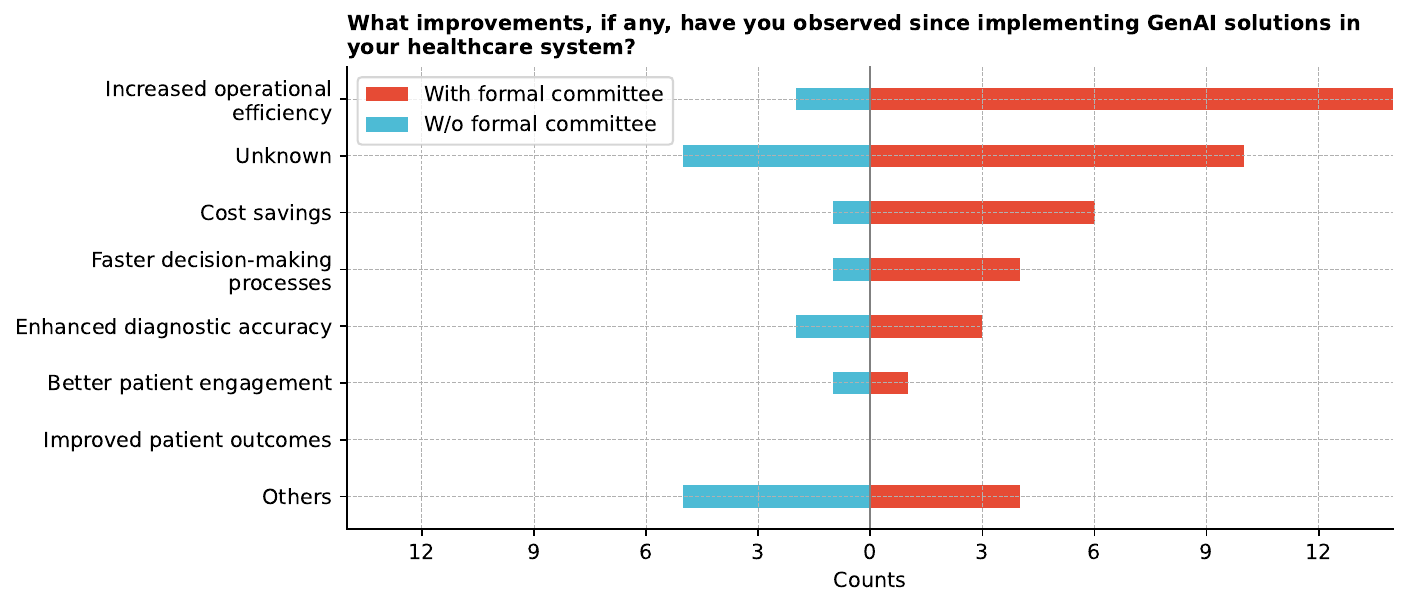}
    \caption{What improvements, if any, have you observed since implementing Generative AI (GenAI) solutions in your healthcare institution?}
    \label{fig:q23}
\end{figure}

Further, respondents reported increased operational efficiency (44.4\%) as the most commonly observed improvement, with faster decision-making processes noted by 13.9\% (\textbf{Figure \ref{fig:q23}}). However, none reported improved patient outcomes. Other reported improvements included increased patient satisfaction and enhanced research capacity, although some noted it was too early to prove such benefits. Significant differences among these improvements were observed (Q = 38.9, p \textless{} 0.0001), particularly between better patient engagement and improved patient outcomes (corrected p = 0.0026) (See \textbf{Supplementary Table \ref{sup tab:LLM improvement}}).

\begin{figure}
    \centering
    \includegraphics[width=\linewidth]{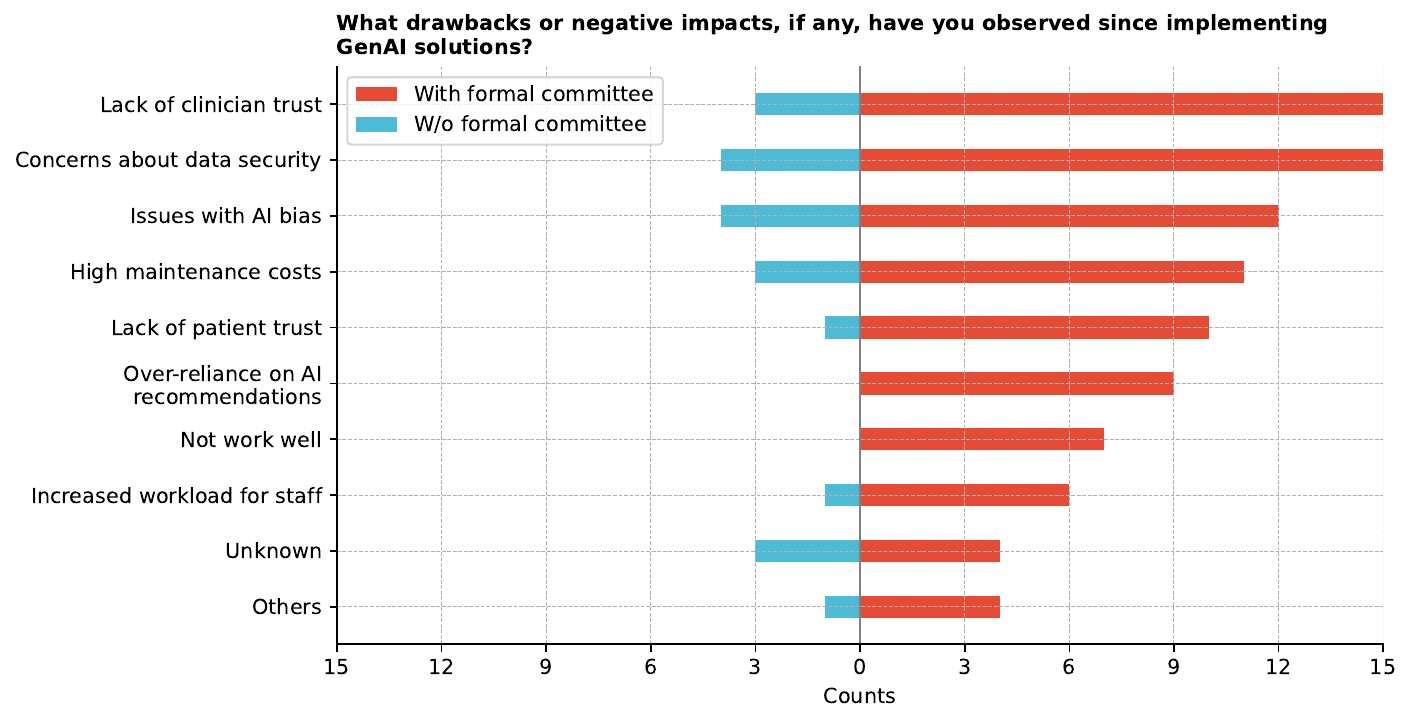}
    \caption{What drawbacks or negative impacts, if any, have you observed since implementing GenAI solutions?}
    \label{fig:q24}
\end{figure}

Regarding GenAI implementation concerns (\textbf{Figure \ref{fig:q24}}), data security was identified as a major issue by 52.78\% of respondents, followed by a lack of clinician trust (50.0\%) and AI bias (44.44\%). Cochran's Q Test confirmed variability in these concerns (Q = 33.3, p \textless{} 0.001). Other challenges included the time required to train models, lack of validation tools, inadequate provider training, and concerns about organizational trust. Some respondents also noted that their observations were based on internal experiences, with no implementations yet in production.
\subsection{Enhancement Strategies}\label{enhancement-strategies}

\begin{figure}
    \centering
    \includegraphics[width=\linewidth]{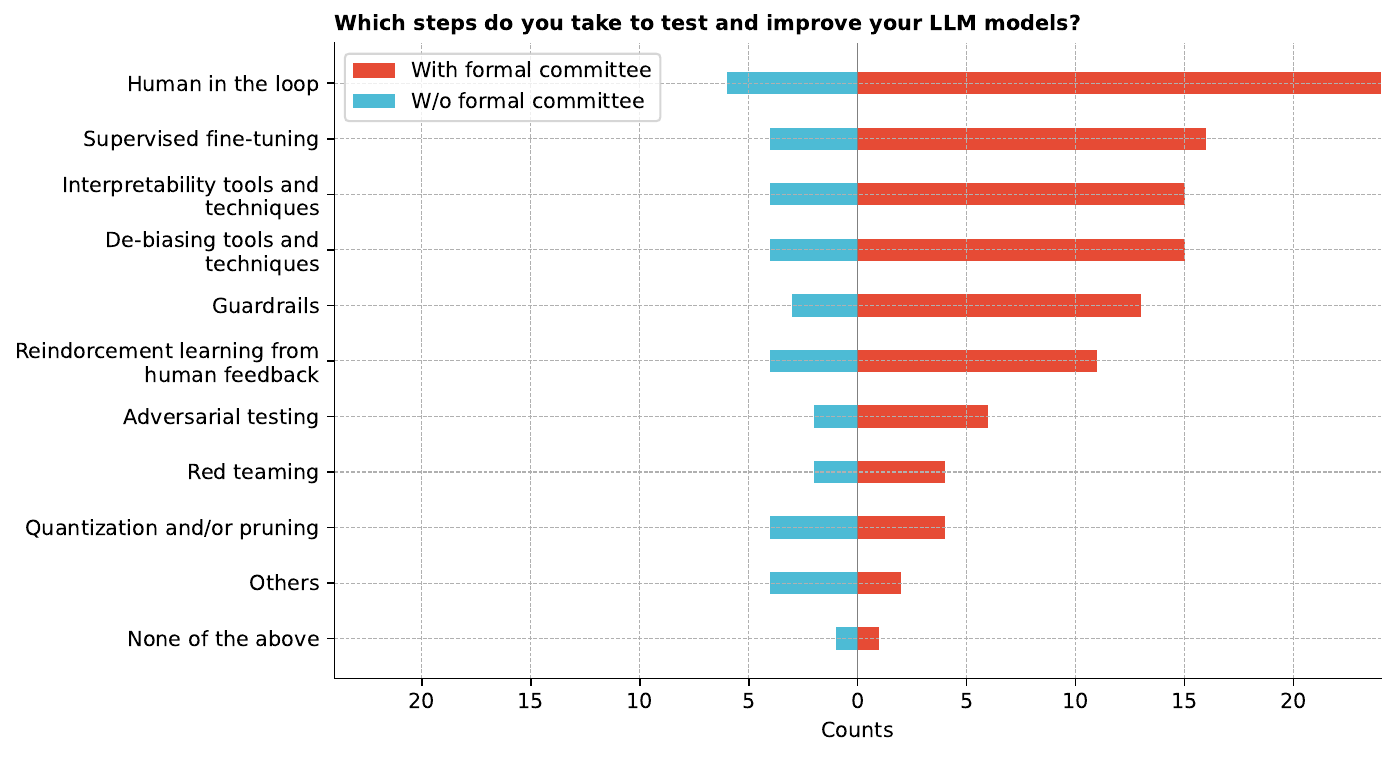}
    \caption{Which steps do you take to test and improve your LLM models?}
    \label{fig:q27}
\end{figure}

Respondents identified several strategies for testing and improving LLMs in healthcare, with human-in-the-loop being the most common (83.3\%, \textbf{Figure \ref{fig:q27}}). Significant differences were noted between human-in-the-loop and methods like quantization and pruning and Reinforcement Learning with human feedback (RLHF)\cite{Stiennon2020-or} (corrected p \textless{} 0.005) (See \textbf{Supplementary Table \ref{sup tab:LLM RLHF}}). Significant differences were found between adversarial testing\cite{Hao2023-ra} and human-in-the-loop (corrected p \textless{} 0.0001) and guardrails and human-in-the-loop (corrected p = 0.0067) (See \textbf{Supplementary Table \ref{sup tab:LLM RLHF}}).

\begin{figure}
    \centering
    \includegraphics[width=\linewidth]{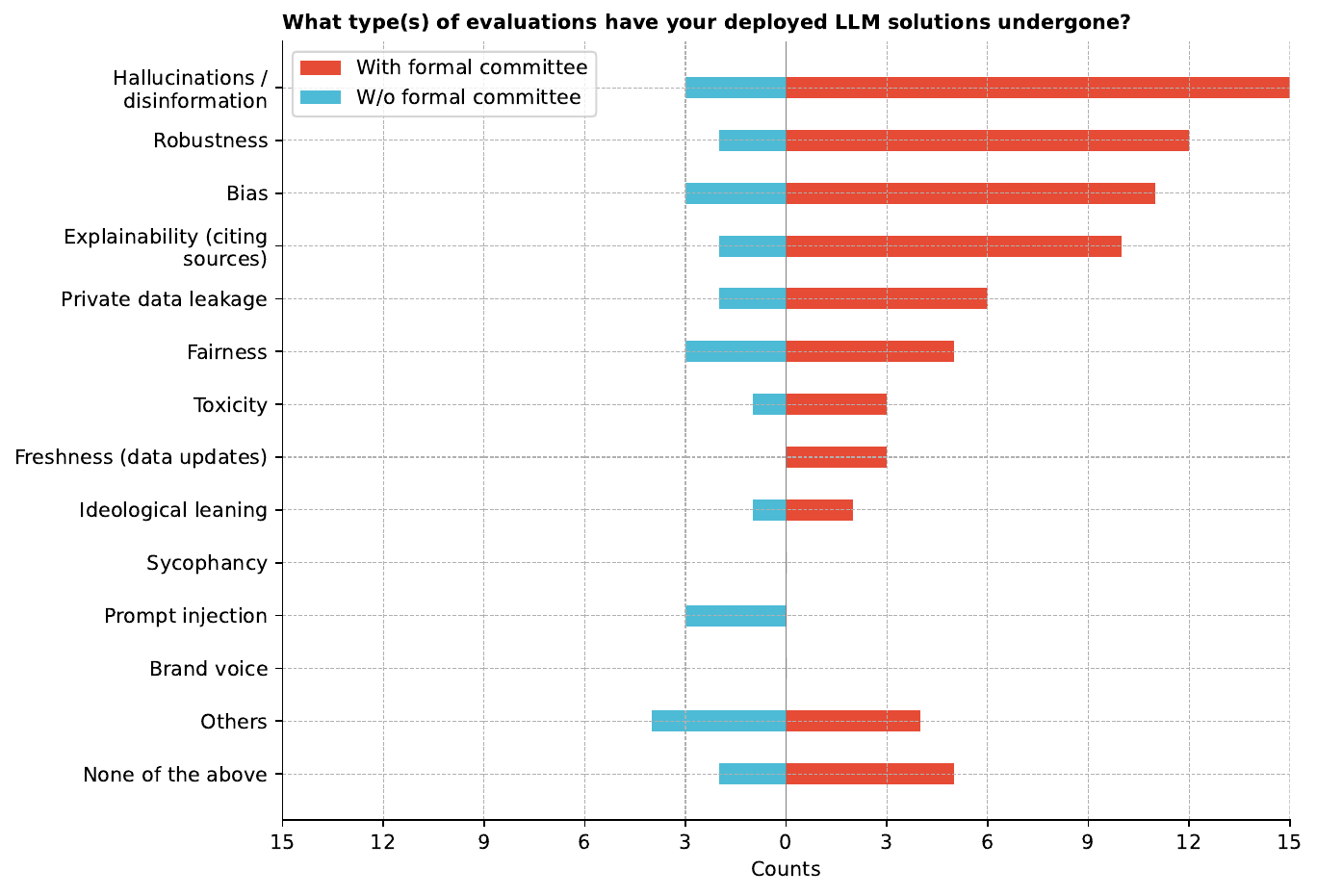}
    \caption{What type(s) of evaluations have your deployed LLM solutions undergone?}
    \label{fig:q28}
\end{figure}

In evaluating deployed LLMs (\textbf{Figure \ref{fig:q28}}), the most common assessments focused on hallucinations or disinformation (50.0\%) and robustness (38.9\%). However, 19.4\% (7/36) of respondents indicated no evaluations had been conducted. Cochran's Q Test revealed significant variation in the importance of these evaluations (Q = 77.1, p \textless{} 0.0001), with post-hoc analysis showing significant differences between explainability and prompt injection (i.e., a technique where specific prompts or questions are used to trick the GenAI into bypassing its specified restrictions, revealing weaknesses in how it understands and responds to information), and between fairness versus ideological leaning and prompt injection (corrected p = 0.0040) (See \textbf{Supplementary Table \ref{sup tab:LLM evaluation}}).

\begin{figure}
    \centering
    \includegraphics[width=\linewidth]{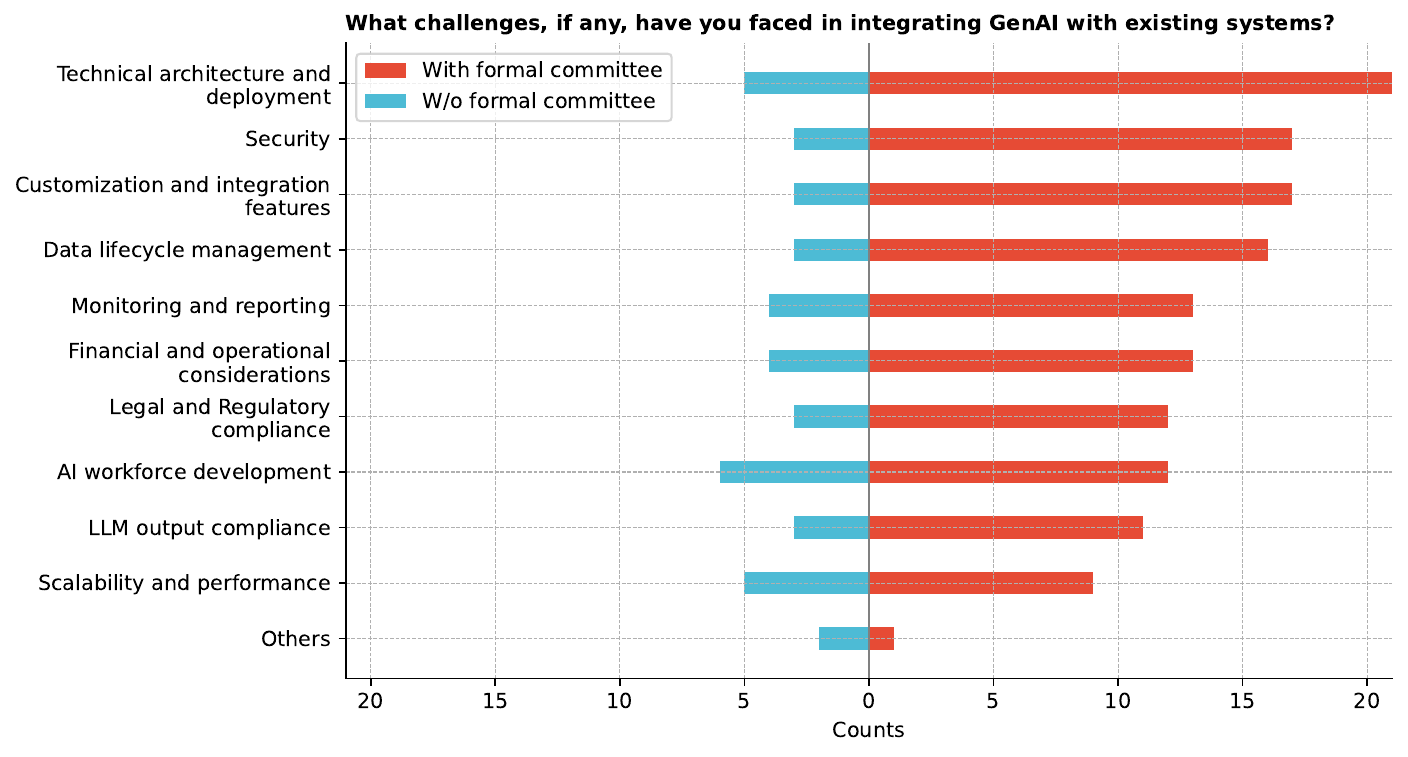}
    \caption{What challenges, if any, have you faced in integrating GenAI with existing systems?}
    \label{fig:q29}
\end{figure}

Integrating GenAI into healthcare presents several challenges (\textbf{Figure \ref{fig:q29}}), with technical architecture and deployment cited most frequently (72.2\%). Interestingly, AI workforce development is the most common challenge for institutions without a formal committee. Data lifecycle management was noted as a critical limitation by 52.8\% (19/36) of respondents. Challenges often overlap, with technical architecture and deployment closely linked to security, scalability, and regulatory compliance issues. Additional gaps were also highlighted, such as the absence of a training plan and a limited workforce. Significant variability was observed (Q = 45.4, p \textless{} 0.0001), with post-hoc analysis indicating that technical architecture and deployment were more prevalent than LLM output compliance (i.e., the trustworthiness of the LLM output) and scalability and performance (corrected p = 0.0269) (See \textbf{Supplementary Table \ref{sup tab:LLM challenge}}).

\subsection{Additional Insights into GenAI Integration}\label{additional-insights-into-genai-integration}

Nine respondents provided additional insights into the complexities of integrating GenAI into healthcare. They emphasized the challenges posed by the rapid pace of technological change, which complicates long-term investment and integration decisions. Organizational approaches to GenAI vary; some institutions aggressively pursue it, while others have yet to implement it on a broader scale despite individual use. The integration of GenAI has improved collaboration between researchers, physicians, and administrators, but slow decision-making and a significant gap in AI workforce skills remain critical issues. The evolving nature of AI initiatives makes it difficult to fully capture current practices, highlighting the need for a comprehensive approach that addresses technological, organizational, and workforce challenges.

\section{DISCUSSION}\label{discussion}

This study provides a snapshot of GenAI integration within CTSA institutions, focusing on key stakeholders, governance structures, ethical considerations, and associated challenges and opportunities. \textbf{Table \ref{tab:findings}} summarizes the key recommendations from the findings. Senior leaders, IT staff, and researchers are central to GenAI integration, with significant involvement from cross-functional committees highlighting the multidisciplinary collaboration required for effective implementation. However, findings suggest minimal involvement of nurses, patients, and community representatives in the current GenAI implementation decision-making process, which raises concerns about inclusiveness, which is essential to aligning technologies with the needs of all stakeholders.\cite{Scott2021-iu, Sujan2022-wm}
Most institutions adopt a centralized, top-down governance structure, streamlining decision-making but potentially limiting flexibility for departmental needs.\cite{Argyres1995-dw}
While formal committees or task forces suggest emerging governance frameworks, the variability across institutions indicates that best practices are still evolving.

\begin{table}[]
\caption{Summary of Key Findings and Recommendations for GenAI Implementation in Healthcare.}
\label{tab:findings}
\small
\begin{tabularx}{\textwidth}{>{\raggedright\arraybackslash}p{10em}X}
\toprule
\textbf{Key Finding} & \textbf{Recommendation} \\
\midrule
Stakeholder Involvement & Involve senior leaders, IT staff, researchers, clinicians, and patients to ensure a representative and effective decision-making process. \\
\rowcolor[gray]{.9}
Governance Structure & Establish formal GenAI governance committees to ensure structured oversight. \\
Decision-Making & Cross-functional committees should lead decision-making for GenAI adoption, balancing stakeholder involvement. \\
\rowcolor[gray]{.9}
Popular Enhancement Strategies & Use human-in-the-loop and supervised fine-tuning as primary enhancement strategies for LLM models. \\
Cloud Architecture Preferences & Prefer private cloud or on-premises hosting to maintain control over security, scalability, and regulatory compliance in GenAI deployment. \\
\rowcolor[gray]{.9}
Ethical Considerations & Prioritize bias and fairness, patient privacy, and data security when integrating GenAI into healthcare institutions. \\
Budget Allocation & Encourage institutions to establish systematic funding mechanisms for GenAI projects to support long-term investments. \\
\rowcolor[gray]{.9}
LLM Usage & Adopt a combination of open and proprietary LLMs, depending on the technical and scalability requirements of the institution. \\
Workforce Training & Implement comprehensive training programs to enhance GenAI literacy and bridge skill gaps within the healthcare workforce. \\
\rowcolor[gray]{.9}
Projected Impact and Improvements & Focus on operational efficiency and decision-making speed while addressing the gap in direct improvements to patient outcomes. \\
\bottomrule
\end{tabularx}
\end{table}

According to the respondents, ethical and regulatory oversight of GenAI implementation varies across institutions, with some involvement from federal agencies, IRBs, and ethics committees. Prioritization of ethical considerations such as patient privacy, data security, and fairness in AI algorithms reflects the awareness of the significant challenges in deploying GenAI in healthcare. Our findings also reveal variability in the reported involvement of regulatory bodies, with less frequent mentions of engagement from local health authorities. However, we did not collect detailed information on the specific roles of these agencies or distinguish between different types of regulatory engagement. This limitation suggests a need for more explicit and consistent oversight frameworks to address the unique risks associated with GenAI. Despite these gaps, this study emphasizes the importance of developing comprehensive policies and guidelines to navigate the ethical landscape of GenAI technologies in healthcare.

Collaboration with vendors is common among CTSA institutions, with partnerships reported with major technology providers like Microsoft Azure AI, Amazon Web Services, Oracle, and Epic Systems. However, the variability in the extent of these collaborations and the need for comprehensive information on enterprise-wide vendor engagements suggest challenges in coordinating AI implementation efforts across institutions. Further, the ad-hoc funding allocation for GenAI projects indicates that AI integration is still in its infancy, with institutions likely testing the waters before committing to substantial investments. Implementing LLMs in healthcare settings presents significant challenges, particularly in technical architecture, deployment, customization, and security, requiring a comprehensive and coordinated approach across departments for successful integration.\cite{Denecke2024-pf}

To evaluate their GenAI technologies, some institutions are using strategies like human-in-the-loop oversight, supervised fine-tuning, and interpretability tools to enhance GenAI transparency and reliability while also employing de-biasing techniques to mitigate biases, ensuring that GenAI outputs are continuously monitored and refined by human experts.\cite{Bakken2023-xf, Mosqueira-Rey2023-zr}
Evaluation practices emphasize robustness and accuracy, with assessments for hallucinations, disinformation, and bias crucial to ascertaining GenAI systems function effectively in real-world healthcare settings.\cite{Williamson2024-se, Menz2024-fp}
However, some institutions' lack of comprehensive evaluations suggests the early stages of LLM adoption and potential shortcomings in initial adoption, highlighting the need to improve their resources or expertise before widespread adoption.

The respondents are optimistic about the projected impact of LLMs on healthcare, particularly in areas like medical text summarization, query interfaces, and information extraction, which are expected to streamline workflows, enhance information access, and improve documentation efficiency.\cite{Tripathi2024-kb, Gebreab2024-xn}
However, the gap between anticipated benefits and actual outcomes, such as the limited direct improvements in patient outcomes, highlights ongoing challenges. This discrepancy emphasizes the need for a focused evaluation of how GenAI tools can directly impact patient health and care quality. Emerging LLM applications in medical education, decentralized trials, and digital twin technologies suggest an expanding scope for these tools. While their impact in specialized domains like drug development remains uncertain, recent evidence points to promising advancements that could enhance the utility of LLMs in this area.\cite{Yang2024-zb}
Despite the enthusiasm, significant concerns about data security, clinician trust, high maintenance costs, AI bias, and lack of patient trust complicate LLM integration into healthcare institutions.

Integrating LLMs into healthcare institutions is further complicated by high maintenance costs, AI bias, and lack of patient trust. Evaluations within institutions prioritize accuracy, reliability, and security, with respondents emphasizing the critical need for dependable and secure AI outputs to maintain trust and patient safety.\cite{Choudhury2024-an}
Legal and reputational risks, along with the need for explainability and transparency, are also highly rated, indicating a significant focus on the ethical and legal implications of AI deployment. However, the importance of these criteria varies, reflecting diverse contexts and priorities across institutions. Despite high expectations for LLMs, the study identified significant roadblocks and considerations for widespread adoption (\textbf{Table \ref{tab:Challenges}}). These challenges underscore the complex landscape where multiple factors must be managed simultaneously.

\begin{table}[]
\caption{Summary of Key Challenges in GenAI Implementation Across CTSA Institutions}
\label{tab:Challenges}
\small
\begin{tabularx}{\textwidth}{>{\raggedright\arraybackslash}p{10em}X}
\toprule
\textbf{Challenge} & \textbf{Description} \\
\midrule
Stakeholder Inclusion & Nurses, patients, and community representatives have limited involvement in the decision-making processes, particularly in institutions without formal committees. \\
\rowcolor[gray]{.9}
Governance Structure & Variability in governance models, with some institutions lacking formal GenAI oversight committees, may impact structured decision-making. \\
Leadership in Decision-Making & Institutions without formal committees rely more on clinical leadership rather than cross-functional committees, potentially affecting the balance of stakeholder input. \\
\rowcolor[gray]{.9}
Ethical Oversight & Varying degree of involvement of ethicists and ethics committees can create gaps and disparity in fairness, privacy, and data security in the broad scientific community for clinical and translational science. \\
Workforce Readiness & Variability in workforce familiarity with LLMs, with some institutions having insufficient training and preparedness for GenAI integration. \\
\rowcolor[gray]{.9}
Training and Skill Gaps & Significant gaps in formal GenAI training plans, with many institutions struggling to build internal capabilities to manage GenAI tools effectively. \\
Technical Integration & Difficulties in integrating GenAI into existing systems, with mixed responses about how well these technologies integrate into current workflows. \\
\rowcolor[gray]{.9}
Funding and Resources & Many institutions rely on ad-hoc funding mechanisms for GenAI projects, creating uncertainty in long-term resource allocation and support for AI initiatives. \\
Vendor Collaboration & Limited transparency and variability in vendor collaborations, with some institutions facing challenges coordinating enterprise-wide AI implementation. \\
\rowcolor[gray]{.9}
Data Security and Trust & Major concerns regarding the security of GenAI systems and lack of clinician trust, particularly in institutions without formal governance structures. \\
AI Bias and Mistrust & Concerns about bias in GenAI outputs and mistrust from clinicians and patients could affect the adoption and effective use of GenAI technologies. \\
\rowcolor[gray]{.9}
Compliance and Legal Risks & Regulatory compliance and accuracy are major concerns, with institutions needing to navigate legal and reputational risks associated with GenAI deployment. \\
\bottomrule
\end{tabularx}
\end{table}

Further, the study reveals that most institutions are still in the experimentation phase of GenAI adoption, exploring the technology's potential and building the necessary skills for its practical adoption. Mixed levels of familiarity with LLMs among the workforce and stakeholders indicate a significant need for further AI workforce training and clinician engagement to enhance GenAI literacy, ensuring that key stakeholders can manage GenAI effectively. Without proper training, healthcare professionals may struggle to fully leverage these tools, potentially leading to inefficiencies, errors, or privacy or security violations (e.g., inappropriately uploading data).\cite{Hazarika2020-sr, Charow2021-hf}
Previous work suggests a multifaceted and multi-sectorial approach to address these gaps and facilitate knowledge sharing, including implementing structured training programs, offering hands-on workshops, developing mentorship opportunities, and partnering with vendors to provide tailored training specific to the healthcare setting.\cite{frehywot2023equitable-r} This opens the possibility that NCATS and other NIH institutes may want to consider collaborative initiatives to address the questions raised in this research. Additionally, the CTSA network\textquotesingle s emphasis on knowledge sharing could facilitate smoother GenAI adoption across institutions,\cite{Fishman2024-um} particularly for late adopters. By encouraging the dissemination of best practices and lessons learned from early adopters,\cite{Escobar-Rodriguez2014-oo} the CTSA network can help institutions with fewer resources or those facing governance challenges navigate the complexities of GenAI implementation more efficiently.

The study has limitations, including variability in respondents' knowledge and the evolving nature of GenAI practices, which may not capture ongoing progress or changes beyond the survey period. Additionally, the reliance on responses from senior leaders, who may not have full visibility into all aspects of GenAI integration within their institutions, introduces the risk of misreporting or incomplete information. The focus on CTSA institutions may limit the generalizability of the findings to other healthcare organizations, particularly for institutions with fewer resources where these implementation and governance challenges may be especially difficult to address. The survey also did not distinguish between live GenAI systems and those still in development, which limits our ability to assess the operational readiness and deployment status of these tools fully across institutions. Additionally, reliance on self-reported data introduces possible biases.

In conclusion, the study highlights the complex and evolving landscape of GenAI integration in CTSA institutions. By identifying successful strategies and highlighting areas for improvement, this research provides an actionable roadmap for institutions seeking to navigate the complexities of AI integration in healthcare to ensure ethical, equitable, and effective implementation, ultimately contributing to advancing patient care and the broader goals of precision medicine.

\section{METHODS}\label{methods}

\subsection{Study Design}\label{study-design}

This study uses an online survey to conduct an environmental scan of GenAI infrastructure within CTSA institutions through multiple choice, ranking, rating, and open-ended questions to understand GenAI integration, including stakeholder roles, governance structures, and ethical considerations.

\subsection{Survey Instrument Development}\label{survey-instrument-development}

The survey, administered through the Qualtrics platform (Qualtrics, Provo, UT), was intended to take approximately 15 minutes to complete. Initially developed through a comprehensive review of current literature on AI in healthcare, the survey covered topics such as stakeholder roles, governance structures, ethical considerations, AI adoption stages, budget trends, and LLM usage. The survey was reviewed by experts (SL, BM, KN, WP, RZ, YZ) in health informatics, clinical practice, ethics, and law, who provided feedback that informed revisions to improve clarity and comprehensiveness. A small group piloted the final version to identify any remaining issues. The survey questions are available in the \textbf{Supplementary File \ref{sec:questionnaire}}.

\subsection{Participant Recruitment}\label{participant-recruitment}

Participants were recruited in July 2024 through targeted outreach to key stakeholders at CTSA sites using purposive and snowball sampling.\cite{Biernacki1981-bm}
Email invitations were sent to senior leaders involved in GenAI implementation and decision-making within the CTSA network (\href{https://ccos-cc.ctsa.io/resources/hub-directory}), with follow-up reminders to maximize response rates.

\subsection{Data Collection}\label{data-collection}

Data were collected from July to August 2024. CTSA leaders who responded to the initial invitation received a follow-up email with the survey link. A PDF version of the survey was provided to help participants prepare by reviewing questions offline before completing the survey online. Participants could return to the survey if necessary.

\subsection{Data Analysis}\label{data-analysis}

Quantitative data from the survey were analyzed using various methods. Multiple-choice and multiple-answer questions were summarized with frequency distributions and percentages. In addition, multiple-answer questions were also analyzed using co-occurrence and pattern analysis to identify common selections and combinations among stakeholder groups. Cochran's Q test identified overall differences among response proportions, with post-hoc analysis using pairwise McNemar tests with Bonferroni corrections.\cite{Stephen2018-ij}
Ranking questions were analyzed by calculating mean ranks, with lower mean ranks indicating higher importance. Likert-scale items were summarized using measures of central tendency and dispersion, with an ANOVA test to check for significant differences in ratings across different use cases, followed by Tukey's HSD test for post-hoc pairwise comparisons while controlling for the family-wise error rate.\cite{Mircioiu2017-jt}

Qualitative data from open-ended survey questions was analyzed using thematic analysis.\cite{Nowell2017-on} This process involved coding the data to identify common themes and patterns. Two researchers (BI, ZX) independently coded the data, and a third researcher (YP) resolved disagreements through consensus.

\section*{DATA AVAILABILITY}\label{data-availability}

The data supporting the findings of this study, with identifying information removed to ensure confidentiality, are available from the corresponding author upon reasonable request. The authors declare that all other data supporting the findings of this study are available within the paper and its supplementary information files.

\section*{ACKNOWLEDGMENTS}

This work was supported by the National Center for Advancing Translational Sciences (NCATS) of the National Institutes of Health (NIH) under grant numbers UL1TR002384, UM1TR004789, UL1TR001412, UL1TR001449, ULTR002345, UM1TR004404, UL1TR001866, UM1TR004909, and UL1TR001873; National Library of Medicine (NLM) of NIH under grant numbers T15LM007079, T15LM012495, R25LM014213; and the National Institute on Alcohol Abuse and Alcoholism of NIH grant numbers R21AA026954 and R33AA0226954. This study was also funded in part by the Department of Veterans Affairs and NIH Intramural Research Program. The content is solely the responsibility of the authors and does not necessarily represent the official views of the NIH.

\section*{AUTHOR CONTRIBUTIONS}\label{author-contributions}

B. Idnay (Methodology, Software, Validation, Formal analysis, Writing), Z. Xu (Data Curation, Formal analysis, Writing), W.G. Adams, M. Adibuzzaman, N.R. Anderson, N. Bahroos, D.S. Bell, C. Bumgardner, T. Campion, M. Castro, P.L. Elkin, J.W. Fan, D.J. Foran, D. Hanauer, M. Hogarth, M. Kandpal, A. Katoch, A. Lai, C.G. Lambert, L. Li, C. Lindsell, J. Liu, Z. Lu, Y. Luo, P. McGarvey, P. Mirhaji, J.D. Osborne, P.A. Harris, F. Prior, N.J. Shaheen, I. Sim, L.R. Waitman (Revision), R.J. Wright, A.H. Zai, K. Zheng (Survey participants, Writing), J.J. Cimino, I.G. Cohen, D. Dorr, T. Ferris, K. Huang, J. Kalpathy-Cramer, N.S. Karnik, E.A. Mendonca, S. Murphy, I. C. Paschalidis, N. Shara, U. Tachinardi (Writing), S.S.J. Lee, B. Malin, K. Natarajan, W.N. Price II, R. Zhang, Y. Zhang (Design of questions, Writing), H. Xu, J. Bian, C. Weng, Y. Peng (Supervision, Conceptualization, Writing, Funding acquisition).

\section*{COMPETING INTERESTS}\label{competing-interests}

I.G. Cohen is a member of the Bayer Bioethics Council, the Chair of the ethics advisory board for Illumina, and an advisor for World Class Health. He was also compensated for speaking at events organized by Philips with the Washington Post, by the Doctors Company, and attending the Transformational Therapeutics Leadership Forum organized by Galen Atlantica. He has been retained as an expert in health privacy, gender-affirming care, and reproductive technology lawsuits.

\bibliographystyle{medline}
\bibliography{ref}

\newpage
\appendix
\setcounter{table}{0}
\setcounter{figure}{0}
\renewcommand\figurename{Supplementary Figure} 
\renewcommand\tablename{Supplementary Table}

\begin{center}
    \large \textbf{Supplementary materials}
\end{center}

\section{Questionnaire}
\label{sec:questionnaire}

\uline{Organization information}

Organization name *:

Q1. Do you represent: (Select all that apply)
\begin{compactitem}
\item
  a CTSA (academic institution)
\item
  A hospital or healthcare system site
\item
  Other (Please specify)
\end{compactitem}

\uline{Stakeholder Identification and Roles}

Q2. Which stakeholder groups are involved in your organization\textquotesingle s decision-making and implementation of Generative AI? (Select all that apply)

\begin{compactitem}
\item
  Hospital administration: Senior leaders (e.g., CEO, CMO, CIO, CMIO/CHIO, or CRIO)
\item
  Hospital administration: Departmental leaders (e.g., medical directors, department chairs, or division chiefs)
\item
  Hospital administration: Business unit leaders (e.g., nursing supervisors)
\item
  Physicians
\item
  Nurses
\item
  Researchers
\item
  IT staff (Data scientists, Data engineers, Health IT)
\item
  Regulatory bodies (Legal, IRB)
\item
  Policy groups (Local governance committees such as DEI committee and ethics committee)
\item
  Patients
\item
  Patient and community representatives
\item
  Others (Please specify) \makebox[1cm]{\hrulefill}
\end{compactitem}

Q3. Who leads the decision-making process for implementing Generative AI applications in your organization? (Select all that apply)

\begin{compactitem}
\item
  Hospital Administration with clinical expertise
\item
  Cross-functional Committee (i.e., individuals with different backgrounds, roles, and expertise)
\item
  Hospital Administration without clinical expertise
\item
  IT Department
\item
  Regulatory bodies
\item
  Others (Please specify) \makebox[1cm]{\hrulefill}
\end{compactitem}

\uline{Decision-Making and Governance Structure}

Q4. How are decisions regarding adopting Generative AI made in your healthcare system? (Select all that apply)

\begin{compactitem}
\item
  Centralized (Top-down)
\item
  Decentralized (Bottom-up; Individual departments independently decide to implement Generative AI based on their specific needs and experiences)
\item
  Collaborative (Involving multiple departments)
\item
  Others (Please specify) \makebox[1cm]{\hrulefill}
\item
  I don't know
\end{compactitem}

Q5. Are there any formal committees or task forces established to oversee the deployment and governance of Generative AI in your organization?

\begin{compactitem}
\item
  Yes\\
  If yes, please describe the expertise and structure of the governance in your organization. \makebox[1cm]{\hrulefill}
\item
  No
\item
  I don't know
\end{compactitem}

\uline{Regulatory and Ethical Considerations}

Q6. Which regulatory bodies are involved in overseeing the deployment of Generative AI in your organization? (Select all that apply)

\begin{compactitem}
\item
  Federal agencies (e.g., FDA, HHS, DOC)
\item
  State agencies
\item
  Local Health Authorities
\item
  Others
\end{compactitem}

Q7. Do you have an ethicist or an ethics committee involved in the decision-making process for implementing Generative AI technologies in your organization?

\begin{compactitem}
\item
  Yes, an ethicist is involved
\item
  Yes, an ethics committee is involved
\item
  No, neither an ethicist nor an ethics committee is involved
\item
  I don't know
\end{compactitem}

Q8. Please rank the following ethical considerations from most important (1) to least important (6) when decision-makers are deciding to implement Generative AI technologies:

\begin{compactitem}
\item
  Bias and fairness of the generative AI algorithms
\item
  Robustness of the generative AI algorithms
\item
  Informed consent from patients
\item
  Patient privacy and data security
\item
  Transparency and explainability of AI decisions made by the algorithms
\item
  Others (Please specify)
\end{compactitem}

\uline{Stage of Adoption}

Q9. What is the stage of generative AI adoption in your organization?

\begin{compactitem}
\item
  Aware: Not actively considering Generative AI as a solution.
\item
  Experimenting: Exploring the potential of Generative AI, building skills and expertise to improve communication, and identifying areas where Generative AI is adding value to the business
\item
  Optimizing: Improving the AI literacy of your entire workforce, not just certain individuals or teams. A first solution is now running in production, as a patient-facing or mission-critical system.
\item
  Standardizing: Ensuring standardized usage of gen AI across the business. Multiple solutions are now running in production.
\item
  Transforming: maximizing the benefits in productivity, business communication, customer satisfaction, and the bottom line.
\end{compactitem}

Q10. How well do Generative AI applications integrate with your existing systems and workflows? For CTSA, it could be OHDSI, ACT, etc. networks. For a healthcare system, it could be clinical workflow (e.g., EPIC).

\begin{compactitem}
\item
  Very poorly
\item
  Poorly
\item
  Neutral
\item
  Well
\item
  Very well
\end{compactitem}

Q11. How familiar are members of the workforce with the use of LLMs in your organization?

\begin{compactitem}
\item
  Not at all familiar
\item
  Sightly familiar
\item
  Somewhat familiar
\item
  Moderately familiar
\item
  Extremely familiar
\end{compactitem}

Q12. Has the workforce received any specific training related to LLMs in your organization?

\begin{compactitem}
\item
  No, and not considered
\item
  No, but considered
\item
  Yes
\end{compactitem}

Q13. How desirable is it for the workforce to receive further LLM training?

\begin{compactitem}
\item
  Very undesirable
\item
  Undesirable
\item
  Neutral
\item
  Desirable
\item
  Very desirable
\end{compactitem}

Q14. Is your institution working with a vendor to implement solutions?

\begin{compactitem}
\item
  No, and not considered
\item
  No, but considered
\item
  Yes. Please indicate how many. \makebox[1cm]{\hrulefill}
\end{compactitem}

\uline{Budget Trends}

Q15. Have funds been allocated for generative AI projects

\begin{compactitem}
\item
  No, and not considered
\item
  No, but considered
\item
  Yes, systematically
\item
  Yes, ad-hoc
\end{compactitem}

Q16. Compared to 2021, the budget allocated to generative AI projects in your organization has:

\begin{compactitem}
\item
  Increased by more than 300\%
\item
  Increased by 100-300\%
\item
  Increased by 50-100\%
\item
  Increased by 10-50\%
\item
  Remained roughly the same (±10\%)
\item
  Unknown
\end{compactitem}

\uline{Current LLM usage}

Q17. Which of the following types of LLM are you currently using? (Select all that apply)

\begin{compactitem}
\item
  Untuned open LLMs (e.g., Llama, Falcon, and Mistral)
\item
  Fine-tuned open LLMs (e.g., Llama, Falcon, and Mistral)
\item
  Untuned proprietary LLMs (e.g., ChatGPT)
\item
  Fine-tuned proprietary LLMs (e.g., ChatGPT)
\item
  Combination of both open and proprietary LLMs
\item
  Uncertain
\item
  We are not using an LLM
\end{compactitem}

Q18. What AI deployment options does your organization currently use?

\begin{compactitem}
\item
  Cloud computing: Public cloud
\item
  Cloud computing: Private cloud
\item
  Cloud computing: Hybrid cloud
\item
  On premises (self-hosting)
\item
  Hybrid
\item
  Others (Please specify) \makebox[1cm]{\hrulefill}
\end{compactitem}

Q19. You indicated that your organization is using open LLMs. What factors influenced your decision to develop internally? (Select all that apply)

\begin{compactitem}
\item
  Technical architecture and deployment (e.g., LLM type, deployment options, infrastructure needs, data handling, compatibility)
\item
  Data lifecycle management (e.g., data preprocessing, customer data use, data sharing, privacy, dedication features, sensitive information)
\item
  Customization and integration features (e.g., proprietary datasets, third-party integrations)
\item
  Security
\item
  Scalability and performance
\item
  LLM output compliance (e.g., bias, output filtering, diversity)
\item
  Legal compliance (e.g., IP considerations, ownership)
\item
  Regulatory compliance (e.g., consent, controls)
\item
  Monitoring and reporting (e.g., transparency, accuracy, use feedback)
\item
  Financial and operational considerations (e.g., costs, support, opt-out features)
\item
  AI workforce development
\item
  Clinician and/or patient buy-in
\item
  Others (Please specify) \makebox[1cm]{\hrulefill}
\end{compactitem}

Q20. You indicated that your organization is using proprietary LLMs. What factors influenced your decision to go with commercial solutions? (Select all that apply)

\begin{compactitem}
\item
  Technical architecture and deployment (e.g., LLM type, deployment options, infrastructure needs, data handling, compatibility)
\item
  Data lifecycle management (e.g., data preprocessing, customer data use, data sharing, privacy, dedication features, sensitive information)
\item
  Customization and integration features (e.g., proprietary datasets, third-party integrations)
\item
  Security
\item
  Scalability and performance
\item
  LLM output compliance (e.g., bias, output filtering, diversity)
\item
  Legal compliance (e.g., IP considerations, ownership)
\item
  Regulatory compliance (e.g., consent, controls)
\item
  Monitoring and reporting (e.g., transparency, accuracy, use feedback)
\item
  Financial and operational considerations (e.g., costs, support, opt-out features)
\item
  AI workforce development
\item
  Clinician and/or patient buy-in
\item
  Others (Please specify) \makebox[1cm]{\hrulefill}
\end{compactitem}

Q21. Which of the following use cases are you currently using LLMs for? (Select all that apply)

\begin{compactitem}
\item
  Data abstraction
\item
  Answering patient questions
\item
  Biomedical research
\item
  Clinical coding or chart audit
\item
  De-identification
\item
  Drug development
\item
  Information extraction
\item
  Machine translation
\item
  Medical chatbot
\item
  Medical image analysis
\item
  Medical text summarization
\item
  Natural language query interface
\item
  Scheduling
\item
  Synthetic data generation
\item
  Transcribing medical encounters
\item
  Others (Please specify) \makebox[1cm]{\hrulefill}
\end{compactitem}

\uline{Projected impact}

Q22. On a scale of 1 to 5, please rate how much you think LLMs will impact each use case over the next 2-3 years. 1 means very negative, and 5 means very positive.

\begin{tabularx}{\textwidth}{|l|*{5}{>{\centering\arraybackslash}X|}}
\hline
~ & 1 & 2 & 3 & 4 & 5  \\ \hline
Data abstraction & ~ & ~ & ~ & ~ &   \\ \hline
Answering Patient Questions & ~ & ~ & ~ & ~ &   \\ \hline
Biomedical Research & ~ & ~ & ~ & ~ &   \\ \hline
Clinical Coding / Chart Audit & ~ & ~ & ~ & ~ &   \\ \hline
De-identification & ~ & ~ & ~ & ~ &   \\ \hline
Drug Development & ~ & ~ & ~ & ~ &   \\ \hline
Information Extraction / Data & ~ & ~ & ~ & ~ &   \\ \hline
Machine Translation & ~ & ~ & ~ & ~ &   \\ \hline
Medical Chatbot & ~ & ~ & ~ & ~ &   \\ \hline
Medical Image Analysis & ~ & ~ & ~ & ~ &   \\ \hline
Medical Text Summarization & ~ & ~ & ~ & ~ &   \\ \hline
Natural Language Query Interface & ~ & ~ & ~ & ~ &   \\ \hline
Scheduling & ~ & ~ & ~ & ~ &   \\ \hline
Synthetic Data Generation & ~ & ~ & ~ & ~ &   \\ \hline
Transcribing Medical Encounters & ~ & ~ & ~ & ~ &   \\ \hline
\end{tabularx}

Q23. What improvements, if any, have you observed since implementing Generative AI solutions in your healthcare system? (Select all that apply)

\begin{compactitem}
\item
  Better patient engagement
\item
  Cost savings
\item
  Enhanced diagnostic accuracy
\item
  Faster decision-making processes
\item
  Improved patient outcomes
\item
  Increased operational efficiency
\item
  Unknown
\item
  Others (Please specify) \makebox[1cm]{\hrulefill}
\end{compactitem}

Q24. What drawbacks or negative impacts, if any, have you observed since implementing Generative AI solutions? (Select all that apply)

\begin{compactitem}
\item
  Concerns about data security
\item
  High maintenance costs
\item
  Increased workload for staff
\item
  Issues with AI bias
\item
  Over-reliance on AI recommendations
\item
  Lack of clinician trust
\item
  Lack of patient trust
\item
  It does not work well
\item
  Unknown
\item
  Others (Please specify) \makebox[1cm]{\hrulefill}
\end{compactitem}

\uline{LLM evaluation (if implementing LLM solutions)}

Q25. On a scale from 1 to 5, please rate the importance of each of the following criteria when evaluating LLMs. 1 means "Not at all Important," and 5 means "Extremely Important."

\begin{tabularx}{\textwidth}{|l|*{5}{>{\centering\arraybackslash}X|}}
\hline
~ & 1 & 2 & 3 & 4 & 5  \\ \hline
Accuracy & ~ & ~ & ~ & ~ &   \\ \hline
Cost & ~ & ~ & ~ & ~ &   \\ \hline
Explainability \& Transparency & ~ & ~ & ~ & ~ &   \\ \hline
Healthcare-Specific Models & ~ & ~ & ~ & ~ &   \\ \hline
Legal \& Reputational Risk & ~ & ~ & ~ & ~ &   \\ \hline
Reproducible \& Consistent Answers & ~ & ~ & ~ & ~ &   \\ \hline
Security \& Privacy Risk & ~ & ~ & ~ & ~ &   \\ \hline
\end{tabularx}

Q26. On a scale of 1 to 5, please rate how significant the following potential limitations or roadblocks are to your roadmap for current generative AI technology, with 1 being not important and 5 being very important.

\begin{tabularx}{\textwidth}{|l|*{5}{>{\centering\arraybackslash}X|}}
\hline
~ & 1 & 2 & 3 & 4 & 5  \\ \hline
Compliance issue & ~ & ~ & ~ & ~ &   \\ \hline
Falls Short on Bias \& Fairness Requirements & ~ & ~ & ~ & ~ &   \\ \hline
Lacks Accuracy & ~ & ~ & ~ & ~ &   \\ \hline
Not Built for Healthcare \& Life Science & ~ & ~ & ~ & ~ &   \\ \hline
Not Tunable for Private Data or Use Cases & ~ & ~ & ~ & ~ &   \\ \hline
Poses Legal, Security, or Reputational Risks & ~ & ~ & ~ & ~ &   \\ \hline
Too Expensive & ~ & ~ & ~ & ~ &   \\ \hline
\end{tabularx}

\uline{Enhancement strategies}

Q27. Which steps do you take to test and improve your LLM models? (Select all that apply)

\begin{compactitem}
\item
  Adversarial testing
\item
  De-biasing tools and techniques
\item
  Guardrails
\item
  Human in the loop
\item
  Interpretability tools and techniques
\item
  Quantization and/or Pruning
\item
  Red Teaming
\item
  Reinforcement Learning from Human Feedback (RLHF)
\item
  Supervised fine-tuning
\item
  Others (Please specify) \makebox[1cm]{\hrulefill}
\item
  None of the above
\end{compactitem}

Q28. What type(s) of evaluations have your deployed LLM solutions undergone? (Select all that apply)

\begin{compactitem}
\item
  Bias
\item
  Brand Voice
\item
  Explainability (citing sources)
\item
  Fairness
\item
  Freshness (data updates)
\item
  Hallucinations / Disinformation
\item
  Ideological Leaning
\item
  Private Data Leakage
\item
  Prompt Injection
\item
  Robustness
\item
  Sycophancy
\item
  Toxicity
\item
  Others (Please specify) \makebox[1cm]{\hrulefill}
\item
  None of the above
\end{compactitem}

Q29. What challenges, if any, have you faced in integrating Generative AI with existing systems? (Select all that apply)

\begin{compactitem}
\item
  Technical architecture and deployment (e.g., LLM type, deployment options, infrastructure needs, data handling, compatibility)
\item
  Data lifecycle management (e.g., data preprocessing, customer data use, data sharing, privacy, dedication features, sensitive information)
\item
  Customization and integration features (e.g., proprietary datasets, third-party integrations)
\item
  Security
\item
  Scalability and performance
\item
  LLM output compliance (e.g., bias, output filtering, diversity)
\item
  Legal and regulatory compliance (e.g., IP considerations, ownership, controls)
\item
  Monitoring and reporting (e.g., transparency, accuracy, use feedback)
\item
  Financial and operational considerations (e.g., costs, support, opt-out features)
\item
  AI workforce development
\item
  Others (Please specify) \makebox[1cm]{\hrulefill}
\end{compactitem}

Q30. Is there anything else you would like to add that was not covered in this survey? Please provide any additional comments, insights, or information that you believe is relevant to understanding the AI infrastructure within your organization.

\makebox[1cm]{\hrulefill}\makebox[1cm]{\hrulefill}\makebox[1cm]{\hrulefill}\_\_

END OF SURVEY

\newpage

{\small

\begin{table}[!ht]
\caption{Results of post-hoc McNemar tests with Bonferroni correction for stakeholder groups involved in the organization's decision-making and implementation of GenAI.} \label{sup tab:stakeholder}
\resizebox{\linewidth}{!}{
\begin{tabular}{r*{11}{>{\raggedleft\arraybackslash}p{5em}}}
\toprule
~ & Senior leaders & Departmental leaders & Business unit leaders & Clinicians & Nurses & Researchers & IT staff & Regulatory bodies & Policy groups & Patients & Patient representatives\\
\midrule
\rowcolor[gray]{.9}
\multicolumn{1}{l}{p-value} &&&&&&&&&&&\\
Departmental leaders & 0.0117 & - & - & - & - & - & - & - & - & - & - \\ 
Business unit leaders & $<$0.0001 & 0.0005 & - & - & - & - & - & - & - & - & - \\ 
Clinicians & 0.0018 & 0.5811 & 0.0127 & - & - & - & - & - & - & - & - \\ 
Nurses & $<$0.0001 & 0.0001 & 0.7266 & 0.0010 & - & - & - & - & - & - & - \\ 
Researchers & 0.1797 & 0.3438 & $<$0.0001 & 0.0391 & $<$0.0001 & - & - & - & - & - & - \\ 
IT staff & 0.2500 & 0.1094 & $<$0.0001 & 0.0117 & $<$0.0001 & 0.7539 & - & - & - & - & - \\ 
Regulatory bodies & 0.0010 & 0.7744 & 0.0042 & 1.0000 & 0.0013 & 0.1460 & 0.0215 & - & - & - & - \\ 
Policy groups & $<$0.0001 & 0.0923 & 0.0654 & 0.3877 & 0.0225 & 0.0034 & 0.0002 & 0.2266 & - & - & - \\ 
Patients & $<$0.0001 & $<$0.0001 & 0.0391 & $<$0.0001 & 0.0625 & $<$0.0001 & $<$0.0001 & $<$0.0001 & 0.0001 & - & - \\ 
Patient representatives & $<$0.0001 & $<$0.0001 & 0.1797 & 0.0009 & 0.4531 & $<$0.0001 & $<$0.0001 & $<$0.0001 & 0.0018 & 0.6250 & ~ \\ 
Others & $<$0.0001 & $<$0.0001 & 0.0654 & 0.0001 & 0.2266 & $<$0.0001 & $<$0.0001 & $<$0.0001 & 0.0013 & 1.0000 & 0.6875 \\  
\midrule
\rowcolor[gray]{.9}
\multicolumn{1}{l}{Corrrected p-value}  &&&&&&&&&&&\\
Departmental leaders & 0.7734 & - & - & - & - & - & - & - & - & - & - \\ 
Business unit leaders & $<$0.0001 & 0.0342 & - & - & - & - & - & - & - & - & - \\ 
Clinicians & 0.1208 & 1.0000 & 0.8399 & - & - & - & - & - & - & - & - \\ 
Nurses & $<$0.0001 & 0.0096 & 1.0000 & 0.0645 & - & - & - & - & - & - & - \\ 
Researchers & 1.0000 & 1.0000 & 0.0005 & 1.0000 & 0.0001 & - & - & - & - & - & - \\ 
IT staff & 1.0000 & 1.0000 & 0.0001 & 0.7734 & $<$0.0001 & 1.0000 & - & - & - & - & - \\ 
Regulatory bodies & 0.0645 & 1.0000 & 0.2759 & 1.0000 & 0.0866 & 1.0000 & 1.0000 & - & - & - & - \\ 
Policy groups & 0.0020 & 1.0000 & 1.0000 & 1.0000 & 1.0000 & 0.2256 & 0.0161 & 1.0000 & - & - & - \\ 
Patients & $<$0.0001 & $<$0.0001 & 1.0000 & 0.0026 & 1.0000 & $<$0.0001 & $<$0.0001 & 0.0003 & 0.0081 & - & - \\ 
Patient representatives & $<$0.0001 & 0.0003 & 1.0000 & 0.0565 & 1.0000 & $<$0.0001 & $<$0.0001 & 0.0010 & 0.1208 & 1.0000 & - \\ 
Others & $<$0.0001 & $<$0.0001 & 1.0000 & 0.0080 & 1.0000 & $<$0.0001 & $<$0.0001 & 0.0003 & 0.0866 & 1.0000 & 1.0000 \\ 
\bottomrule
\end{tabular}
}
\end{table}}

\newpage


\begin{table}[!ht]
\caption{Results of post-hoc McNemar tests with Bonferroni correction for leaders of the decision-making process for implementing GenAI applications in the organization.}
\label{sup tab:decision}
\resizebox{\linewidth}{!}{
\begin{tabular}{r*{5}{>{\raggedleft\arraybackslash}p{5em}}}
\toprule
& Clinical Leadership & Cross-functional Committee & Hospital Administration & IT Department & Regulatory bodies\\ 
\midrule
\rowcolor[gray]{.9}
\multicolumn{6}{l}{p-value}\\
Cross-functional Committee & 0.0127 & - & - & - & - \\ 
Hospital Administration & 0.7266 & 0.0023 & - & - & - \\ 
IT Department & 1.0000 & 0.0075 & 1.0000 & - & - \\ 
Regulatory bodies & 0.0018 & $<$0.0001 & 0.0129 & 0.0074 & - \\ 
Others & 0.0075 & $<$0.0001 & 0.0414 & 0.0192 & 1.0000 \\ 
\midrule
\rowcolor[gray]{.9}
\multicolumn{6}{l}{Corrrected p-value} \\
Cross-functional Committee & 0.1909 & - & - & - & - \\ 
Hospital Administration & 1.0000 & 0.0352 & - & - & - \\ 
IT Department & 1.0000 & 0.1131 & 1.0000 & - & - \\ 
Regulatory bodies & 0.0275 & $<$0.0001 & 0.1941 & 0.1108 & - \\ 
Others & 0.1131 & 0.0005 & 0.6208 & 0.2882 & 1.0000 \\ 
\bottomrule
\end{tabular}
}
\end{table}

\newpage

{\small
\begin{xltabular}{\textwidth}{X}
\caption{Excerpts of governance and leadership structures in GenAI deployment across CTSA institutions.}
\label{sup tab:experts}\\

\toprule
\textbf{Free Text Answers} \\
\midrule
\endfirsthead
{{\tablename\ \thetable{} -- continued from previous page.}} \\
\toprule
\textbf{Free Text Answers} \\
\midrule
\endhead

\midrule 
\multicolumn{1}{r}{{Continued on next page}} \\ 
\endfoot

\bottomrule
\endlastfoot

Clinical, IT, machine learning, for the most part\\
\rowcolor[gray]{.9}
We have multiple governance committees. We have one focused on the university and non-clinical applications. We have a separate joint governance structure that includes representation from the medical school and health system leadership. Both sets of governance includes informatics leadership and those who are conducting research in Generative AI and would be considered subject matter experts. It also has IT representation from the perspective of supporting generative AI infrastructure, access, and policies. Our governance structure also incorporates clinical and/or academic leadership.\\
Health Data Oversight Committee - oversees Data Acesss Committee, Health Analytics Committee, Data Sharing Committee, Advanced Computing Committee. Each of these subcommitees have representation from clinical/research faculty and IT. HDOC itself includes Dean, VD and senior IT/ roles, and report up to the Vice Chancellor for Human Health Sciences. \\
\rowcolor[gray]{.9}
Committee chaired by the Provost with representatives from Clinical Informatics, Chair of Biomedical informatics, Clinical and Research leadership\\
Healthcare system and School of Medicine have an integrated IT governance system that makes decisions about new technology to be adopted.\\
\rowcolor[gray]{.9}
We have a new AI governance group, generative AI is loosely in the scope of this group. \\
Senior leadership establishes policy and has created an generative AI committee and has partnered with Microsoft to allow OpenAI models to be run at [institution]. However, they disallow identified data to be submitted to Open AI LLM endpoints limiting clinical application. The generative AI committee has nobody with deep AI experience (Computer Science PhD in relevant field or any Computer Scientists at all as far as I know).  Many Departments are experimenting with generative AI, writing papers, proposing research, including trying some biomedical and clinical applications. All implementation and experiments are bottom up, but policy and the Microsoft partnership is top-down.\\
\rowcolor[gray]{.9}
The university has a governance committee of Generative AI that CTSA participates.\\
Clinical AI strategic planning group (clinical, admin, data science, IT) sets direction, Responsible AI Oversight Group provides governance\\
\rowcolor[gray]{.9}
We have operational and regulatory leaders from IT, quality analytics, CISO, CMIO, associate dean for informatics, associate dean for compliance\\
It is within scope for our standing Clinical Artificial Intelligence Committee which evaluates AI-based algorithms and interventions\\
\rowcolor[gray]{.9}
IT and CRIO work collaboratively on the governance of Gen AI.\\
AI Governance committee that guides strategy and completes intake for requests.\\
\rowcolor[gray]{.9}
Work in progress - vetting of tools, regulatory review processes\\
Committee on general AI, including generative \\
\rowcolor[gray]{.9}
Institution has established a number of committees governing implementation of AI.  These include the AI Committee Supporting Teaching, Learning, \& Discovery; the AI Risks, Ethics and Policy Committee; and the Institution HS Artificial Intelligence and Machine Learning Governance (AIGOV) Committee.\\
Knowledge rooted in IT, computer science, informatics, clinicians\\
\rowcolor[gray]{.9}
CMIO, clinician champions, regulatory representatives, IT \\
We have an Enterprise AI Translation Advisory Board led by seasoned clinical informaticians, health IT professionals, data scientists, AI researchers, and legal experts that represent organization executives, IT operation \& governance, technology innovation \& translation, health care delivery, as well as ethics and compliance.\\
\rowcolor[gray]{.9}
See prior responses\\
We have University-wide multi-discipinary work group and a Hosptial-based CMIO lead workgroup\\
\rowcolor[gray]{.9}
The committee continues to evolve but includes clinical, EHR, and IT leadership as well as research leadership where necessary.\\
We have a few tri-institutional governance committees that oversee GenAI implementation. We have a clinical group that reviews utility, another committee that evaluates models and another that reviews health equity.\\
\rowcolor[gray]{.9}
There is strategic and operational governance of genAI deployed in the Health system. Operational governance includes clinicians and researchers with genAI expertise. Outside of the health system, operational governance is partially in the IRB and  in IT Governance committees that do not have sufficient expertise. Strategic governance of non-Health system genAI is at the Chancellor's level and not formally chartered.  \\
Institution consists of two separate legal entities, the University and its Health System. These two institutions have separate IT and governance structures. To be able to conduct biomedical research, especially using clinical data, we have a research data governance structure that negotiates the rules from the two systems. This committee is also working on data and infrastructure sharing with our partner institutions (We are a multi-institutional CTSA).\\
\rowcolor[gray]{.9}
Governance Committee on Data, AI and Analytics\\
The Generative AI Healthcare Workgroup is a broad team with representation from across the hospital and including some clinician researchers and clinical informaticists. The workgroup is led by the CMIO.\\
\rowcolor[gray]{.9}
Clinical informatics leaders with representatives from departments/IT
\end{xltabular}}

\newpage


\begin{table}[!ht]
\caption{Results of post-hoc McNemar tests with Bonferroni correction for current use of LLMs.}
\label{sup tab:llm use}
\begin{tabular}{r*{3}{r}}
\toprule
& We are not using an LLM & Combination of both & Open LLMs only \\
\midrule
\rowcolor[gray]{.9}
\multicolumn{4}{l}{P-value} \\
Combination of both & $<$0.0001 & - & - \\ 
Open LLMs only & 0.3750 & 0.0005 & - \\ 
Proprietary LLMs only & 0.0215 & 0.0294 & 0.2668 \\ 
\midrule
\rowcolor[gray]{.9}
\multicolumn{4}{l}{Corrrected p-value} \\
Combination of both & $<$0.0001 & - & - \\ 
Open LLMs only & 1.0000 & 0.0032 & - \\ 
Proprietary LLMs only & 0.1289 & 0.1767 & 1.0000 \\
\bottomrule
\end{tabular}
\end{table}

\newpage

{\small

\begin{table}[!ht]
\caption{Results of post-hoc McNemar tests with Bonferroni correction for factors influenced the decision to develop internally when the organization uses open LLMs} \label{sup tab:open LLM}
\resizebox{\linewidth}{!}{
\begin{tabular}{r*{12}{>{\raggedleft\arraybackslash}p{5em}}}
\toprule
~ & Technical architecture and deployment & Data lifecycle management & Customization and integration features & Security & Scalability and performance & LLM output compliance & Legal compliance & Regulatory compliance & Monitoring and reporting & Financial and operational considerations & AI workforce development & Clinician and/or patient buy-in \\
\midrule
\rowcolor[gray]{.9}
\multicolumn{1}{l}{p-value} &&&&&&&&&&&&\\
Data lifecycle management & 0.1797 & - & - & - & - & - & - & - & - & - & - & - \\ 
Customization and integration features & 0.3438 & 1.0000 & - & - & - & - & - & - & - & - & - & - \\ 
Security & 0.3018 & 1.0000 & 1.0000 & - & - & - & - & - & - & - & - & - \\ 
Scalability and performance & 0.0023 & 0.0963 & 0.0490 & 0.0963 & - & - & - & - & - & - & - & - \\ 
LLM output compliance & 0.0127 & 0.1460 & 0.0923 & 0.1460 & 0.7905 & - & - & - & - & - & - & - \\ 
Legal compliance & 0.0574 & 0.4531 & 0.2891 & 0.4531 & 0.3018 & 0.5078 & - & - & - & - & - & - \\ 
Regulatory compliance & 0.1460 & 1.0000 & 0.7744 & 1.0000 & 0.1435 & 0.2266 & 0.7266 & - & - & - & - & - \\ 
Monitoring and reporting & 0.0213 & 0.2266 & 0.1094 & 0.2266 & 0.5811 & 1.0000 & 0.6875 & 0.2891 & - & - & - & - \\ 
Financial and operational considerations & 0.2266 & 1.0000 & 1.0000 & 1.0000 & 0.0574 & 0.1796 & 0.5811 & 1.0000 & 0.3018 & - & - & - \\ 
AI workforce development & 0.0063 & 0.1797 & 0.1460 & 0.3018 & 0.6072 & 1.0000 & 0.7744 & 0.3438 & 1.0000 & 0.3018 & - & - \\ 
Clinician and/or patient buy-in & $<$0.0001 & 0.0010 & 0.0005 & 0.0074 & 0.5488 & 0.1250 & 0.0215 & 0.0063 & 0.0313 & 0.0034 & 0.0313 & - \\ 
Others & 0.0013 & 0.0784 & 0.0309 & 0.0931 & 1.0000 & 0.6291 & 0.2632 & 0.1153 & 0.4807 & 0.0352 & 0.4807 & 0.7905 \\ 
\midrule
\rowcolor[gray]{.9}
\multicolumn{1}{l}{Corrrected p-value}  &&&&&&&&&&&&\\
Data lifecycle management & 1.0000 & - & - & - & - & - & - & - & - & - & - & - \\ 
Customization and integration features & 1.0000 & 1.0000 & - & - & - & - & - & - & - & - & - & - \\ 
Security & 1.0000 & 1.0000 & 1.0000 & - & - & - & - & - & - & - & - & - \\ 
Scalability and performance & 0.1833 & 1.0000 & 1.0000 & 1.0000 & - & - & - & - & - & - & - & - \\ 
LLM output compliance & 0.9926 & 1.0000 & 1.0000 & 1.0000 & 1.0000 & - & - & - & - & - & - & - \\ 
Legal compliance & 1.0000 & 1.0000 & 1.0000 & 1.0000 & 1.0000 & 1.0000 & - & - & - & - & - & - \\ 
Regulatory compliance & 1.0000 & 1.0000 & 1.0000 & 1.0000 & 1.0000 & 1.0000 & 1.0000 & - & - & - & - & - \\ 
Monitoring and reporting & 1.0000 & 1.0000 & 1.0000 & 1.0000 & 1.0000 & 1.0000 & 1.0000 & 1.0000 & - & - & - & - \\ 
Financial and operational considerations & 1.0000 & 1.0000 & 1.0000 & 1.0000 & 1.0000 & 1.0000 & 1.0000 & 1.0000 & 1.0000 & - & - & - \\ 
AI workforce development & 0.4951 & 1.0000 & 1.0000 & 1.0000 & 1.0000 & 1.0000 & 1.0000 & 1.0000 & 1.0000 & 1.0000 & - & - \\ 
Clinician and/or patient buy-in & 0.0024 & 0.0762 & 0.0381 & 0.5760 & 1.0000 & 1.0000 & 1.0000 & 0.4951 & 1.0000 & 0.2666 & 1.0000 & - \\ 
 Others & 0.1024 & 1.0000 & 1.0000 & 1.0000 & 1.0000 & 1.0000 & 1.0000 & 1.0000 & 1.0000 & 1.0000 & 1.0000 & 1.0000 \\ 
\bottomrule
\end{tabular}
}
\end{table}}

\newpage


\begin{table}[!ht]
\caption{Results of post-hoc McNemar tests with Bonferroni correction for approaches used in GenAI deployment.}
\label{sup tab:factor commercial}
\begin{tabular}{r*{5}{>{\raggedleft\arraybackslash}p{5em}}}
\toprule
~ & Public cloud & Private cloud & Hybrid cloud & On premises & Hybrid \\ 
\midrule
\rowcolor[gray]{.9}
\multicolumn{1}{l}{P-value}  &&&&&\\
Private cloud & 0.0001 & - & - & - & - \\ 
Hybrid cloud & 1.0000 & 0.0026 & - & - & - \\ 
On premises & 0.0004 & 1.0000 & 0.0015 & - & - \\ 
Hybrid & 0.7539 & 0.0043 & 1.0000 & 0.0026 & - \\ 
Others & 0.5488 & $<$0.0001 & 0.3438 & $<$0.0001 & 0.2266 \\ 
\midrule
\rowcolor[gray]{.9}
\multicolumn{1}{l}{Corrrected p-value} &&&&&\\
Private cloud & 0.0022 & - & - & - & - \\ 
Hybrid cloud & 1.0000 & 0.0390 & - & - & - \\ 
On premises & 0.0060 & 1.0000 & 0.0223 & - & - \\ 
Hybrid & 1.0000 & 0.0652 & 1.0000 & 0.0387 & - \\ 
Others & 1.0000 & 0.0010 & 1.0000 & 0.0003 & 1.0000 \\  
\bottomrule
\end{tabular}
\end{table}

\newpage

{\small
\begin{table}[!ht]
\caption{Results of post-hoc McNemar tests with Bonferroni correction for factors that influenced the decision to go with proprietary LLMs.} \label{sup tab:proprietary LLM}
\resizebox{\linewidth}{!}{
\begin{tabular}{r*{12}{>{\raggedleft\arraybackslash}p{5em}}}
\toprule
~ & Technical architecture and deployment & Data lifecycle management & Customization and integration features & Security & Scalability and performance & LLM output compliance & Legal compliance & Regulatory compliance & Monitoring and reporting & Financial and operational considerations & AI workforce development & Clinician and/or patient buy-in \\
\midrule
\rowcolor[gray]{.9}
\multicolumn{1}{l}{p-value} &&&&&&&&&&&&\\
Data lifecycle management & 0.0018 & - & - & - & - & - & - & - & - & - & - & - \\ 
Customization and integration features & 0.0117 & 0.5811 & - & - & - & - & - & - & - & - & - & - \\ 
Security & 0.1796 & 0.0703 & 0.5811 & - & - & - & - & - & - & - & - & - \\ 
Scalability and performance & 0.7905 & 0.0129 & 0.1435 & 0.4240 & - & - & - & - & - & - & - & - \\ 
LLM output compliance & 0.0004 & 0.3438 & 0.0654 & 0.0213 & 0.0005 & - & - & - & - & - & - & - \\ 
Legal compliance & 0.0023 & 1.0000 & 0.4545 & 0.0654 & 0.0192 & 0.5078 & - & - & - & - & - & - \\ 
Regulatory compliance & 0.0490 & 0.5811 & 1.0000 & 0.6072 & 0.1435 & 0.0654 & 0.2891 & - & - & - & - & - \\ 
Monitoring and reporting & 0.0001 & 0.3438 & 0.1185 & 0.0213 & 0.0005 & 1.0000 & 0.5488 & 0.0654 & - & - & - & - \\ 
Financial and operational considerations & 0.2101 & 0.1796 & 0.6072 & 1.0000 & 0.4240 & 0.0063 & 0.1185 & 0.5488 & 0.0309 & - & - & - \\ 
AI workforce development & 0.0001 & 0.3438 & 0.0654 & 0.0129 & 0.0001 & 1.0000 & 0.5488 & 0.0654 & 1.0000 & 0.0020 & - & - \\ 
Clinician and/or patient buy-in & 0.0005 & 0.7539 & 0.3018 & 0.0386 & 0.0075 & 0.7539 & 1.0000 & 0.3018 & 0.7539 & 0.0574 & 0.7539 & - \\ 
Others & 0.0002 & 0.2668 & 0.0574 & 0.0192 & 0.0007 & 1.0000 & 0.3877 & 0.0768 & 1.0000 & 0.0192 & 1.0000 & 0.5078 \\ 
\midrule
\rowcolor[gray]{.9}
\multicolumn{1}{l}{Corrrected p-value}  &&&&&&&&&&&&\\
Data lifecycle management & 0.1428 & - & - & - & - & - & - & - & - & - & - & ~ \\ 
Customization and integration features & 0.9141 & 1.0000 & - & - & - & - & - & - & - & - & - & ~ \\ 
Security & 1.0000 & 1.0000 & 1.0000 & - & - & - & - & - & - & - & - & ~ \\ 
Scalability and performance & 1.0000 & 1.0000 & 1.0000 & 1.0000 & - & - & - & - & - & - & - & ~ \\ 
LLM output compliance & 0.0314 & 1.0000 & 1.0000 & 1.0000 & 0.0405 & - & - & - & - & - & - & ~ \\ 
Legal compliance & 0.1833 & 1.0000 & 1.0000 & 1.0000 & 1.0000 & 1.0000 & - & - & - & - & - & ~ \\ 
Regulatory compliance & 1.0000 & 1.0000 & 1.0000 & 1.0000 & 1.0000 & 1.0000 & 1.0000 & - & - & - & - & ~ \\ 
Monitoring and reporting & 0.0113 & 1.0000 & 1.0000 & 1.0000 & 0.0405 & 1.0000 & 1.0000 & 1.0000 & - & - & - & ~ \\ 
Financial and operational considerations & 1.0000 & 1.0000 & 1.0000 & 1.0000 & 1.0000 & 0.4951 & 1.0000 & 1.0000 & 1.0000 & - & - & ~ \\ 
AI workforce development & 0.0113 & 1.0000 & 1.0000 & 1.0000 & 0.0095 & 1.0000 & 1.0000 & 1.0000 & 1.0000 & 0.1523 & - & ~ \\ 
Clinician and/or patient buy-in & 0.0405 & 1.0000 & 1.0000 & 1.0000 & 0.5880 & 1.0000 & 1.0000 & 1.0000 & 1.0000 & 1.0000 & 1.0000 & ~ \\ 
Others & 0.0173 & 1.0000 & 1.0000 & 1.0000 & 0.0568 & 1.0000 & 1.0000 & 1.0000 & 1.0000 & 1.0000 & 1.0000 & 1.0000 \\
\bottomrule
\end{tabular}
}
\end{table}}

\newpage

{\small

\begin{table}[!ht]
\caption{Results of Co-Occurrence Analysis for factors influenced the decision to develop internally when the organization uses open LLMs.} 
\label{sup tab:factors open}
\resizebox{\linewidth}{!}{
\begin{tabular}{r*{15}{>{\raggedleft\arraybackslash}p{5em}}}
\toprule
- & Data abstraction & Answering patient questions & Biomedical research & Clinical coding or chart audit & De-identification & Drug development & Information extraction & Machine translation & Medical chatbot & Medical image analysis & Medical text summarization & Natural language query interface & Scheduling & Synthetic data generation & Transcribing medical encounters \\ 
\midrule
Answering patient questions & 10 & - & - & - & - & - & - & - & - & - & - & - & - & - & - \\ 
Biomedical research & 17 & 7 & - & - & - & - & - & - & - & - & - & - & - & - & - \\ 
Clinical coding or chart audit & 11 & 5 & 12 & - & - & - & - & - & - & - & - & - & - & - & - \\ 
De-identification & 8 & 3 & 10 & 4 & - & - & - & - & - & - & - & - & - & - & - \\ 
Drug development & 3 & 0 & 4 & 3 & 1 & - & - & - & - & - & - & - & - & - & - \\ 
Information extraction & 16 & 9 & 18 & 10 & 9 & 2 & - & - & - & - & - & - & - & - & - \\ 
Machine translation & 3 & 1 & 3 & 2 & 1 & 0 & 3 & - & - & - & - & - & - & - & - \\ 
Medical chatbot & 10 & 10 & 9 & 7 & 3 & 1 & 10 & 2 & - & - & - & - & - & - & - \\ 
Medical image analysis & 8 & 5 & 11 & 7 & 5 & 4 & 10 & 3 & 6 & - & - & - & - & - & - \\ 
Medical text summarization & 18 & 11 & 18 & 10 & 8 & 3 & 15 & 4 & 10 & 10 & - & - & - & - & - \\ 
Natural language query interface & 12 & 7 & 13 & 6 & 6 & 2 & 11 & 4 & 9 & 8 & 12 & - & - & - & - \\ 
Scheduling & 1 & 1 & 0 & 0 & 0 & 0 & 1 & 0 & 1 & 0 & 1 & 0 & - & - & - \\ 
Synthetic data generation & 7 & 4 & 8 & 4 & 5 & 1 & 9 & 0 & 5 & 7 & 7 & 7 & 0 & - & - \\ 
Transcribing medical encounters & 11 & 9 & 11 & 7 & 6 & 2 & 11 & 2 & 6 & 7 & 13 & 7 & 1 & 4 & - \\ 
Other & 2 & 1 & 2 & 1 & 1 & 0 & 2 & 2 & 1 & 0 & 0 & 0 & 0 & 0 & 2 \\ 
\bottomrule
\end{tabular}
}
\end{table} }

\newpage

{\small
\begin{table}[!ht]
\caption{Results of post-hoc McNemar tests with Bonferroni correction for LLM use cases.} \label{sup tab:LLM use}
\resizebox{\linewidth}{!}{
\begin{tabular}{r*{15}{>{\raggedleft\arraybackslash}p{5em}}}
\toprule
~ & Data abstraction & Answering patient questions & Biomedical research & Clinical coding or chart audit & De-identification & Drug development & Information extraction & Machine translation	& Medical chatbot & Medical image analysis & Medical text summarization & Natural language query interface & Scheduling & Synthetic data generation & Transcribing medical encounters\\
\midrule
\rowcolor[gray]{.9}
\multicolumn{1}{l}{p-value} &&&&&&&&&&&&&&&\\
Answering patient questions & 0.0490 & - & - & - & - & - & - & - & - & - & - & - & - & - & - \\
Biomedical research & 1.0000 & 0.0639 & - & - & - & - & - & - & - & - & - & - & - & - & - \\
Clinical coding or chart audit & 0.0129 & 1.0000 & 0.0034 & - & - & - & - & - & - & - & - & - & - & - & - \\
De-identification & 0.0023 & 0.4807 & 0.0001 & 0.6072 & - & - & - & - & - & - & - & - & - & - & - \\
Drug development & $<$0.0001 & 0.0309 & $<$0.0001 & 0.0117 & 0.1460 & - & - & - & - & - & - & - & - & - & - \\
Information extraction & 0.7744 & 0.1435 & 0.5078 & 0.0574 & 0.0034 & 0.0002 & - & - & - & - & - & - & - & - & - \\
Machine translation & $<$0.0001 & 0.0213 & $<$0.0001 & 0.0225 & 0.1460 & 1.0000 & $<$0.0001 & - & - & - & - & - & - & - & - \\
Medical chatbot & 0.0213 & 1.0000 & 0.0192 & 1.0000 & 0.6291 & 0.0352 & 0.0574 & 0.0225 & - & - & - & - & - & - & - \\
Medical image analysis & 0.0414 & 1.0000 & 0.0074 & 1.0000 & 0.5811 & 0.0039 & 0.0574 & 0.0117 & 1.0000 & - & - & - & - & - & - \\
Medical text summarization & 1.0000 & 0.0213 & 1.0000 & 0.0127 & 0.0013 & $<$0.0001 & 0.6072 & $<$0.0001 & 0.0127 & 0.0127 & - & - & - & - & - \\
Natural language query interface & 0.0574 & 1.0000 & 0.0225 & 0.8036 & 0.2668 & 0.0074 & 0.1796 & 0.0010 & 0.7539 & 0.7744 & 0.0352 & - & - & - & - \\
Scheduling & $<$0.0001 & 0.0002 & $<$0.0001 & 0.0018 & 0.0117 & 0.3750 & $<$0.0001 & 0.3750 & 0.0005 & 0.0018 & $<$0.0001 & 0.0005 & - & - & - \\
Synthetic data generation & 0.0013 & 0.3018 & 0.0003 & 0.4240 & 1.0000 & 0.2266 & 0.0005 & 0.1797 & 0.3877 & 0.2891 & 0.0007 & 0.1094 & 0.0215 & - & - \\
Transcribing medical encounters & 0.2379 & 0.5811 & 0.1671 & 0.4545 & 0.1185 & 0.0023 & 0.4545 & 0.0023 & 0.4807 & 0.4545 & 0.1185 & 0.8145 & $<$0.0001 & 0.0963 & - \\
Other & 0.0003 & 0.0490 & 0.0002 & 0.0768 & 0.2668 & 1.0000 & 0.0009 & 1.0000 & 0.0768 & 0.0963 & 0.0005 & 0.0309 & 0.2188 & 0.4240 & 0.00750 \\
\midrule
\rowcolor[gray]{.9}
\multicolumn{1}{l}{Corrrected p-value}  &&&&&&&&&&&&&&&\\
Answering patient questions & 1.0000 & - & - & - & - & - & - & - & - & - & - & - & - & - & - \\ 
Biomedical research & 1.0000 & 1.0000 & - & - & - & - & - & - & - & - & - & - & - & - & - \\ 
Clinical coding or chart audit & 1.0000 & 1.0000 & 0.4102 & - & - & - & - & - & - & - & - & - & - & - & - \\ 
De-identification & 0.2820 & 1.0000 & 0.0146 & 1.0000 & - & - & - & - & - & - & - & - & - & - & - \\ 
Drug development & 0.0025 & 1.0000 & 0.0002 & 1.0000 & 1.0000 & - & - & - & - & - & - & - & - & - & - \\ 
Information extraction & 1.0000 & 1.0000 & 1.0000 & 1.0000 & 0.4102 & 0.0266 & - & - & - & - & - & - & - & - & - \\ 
Machine translation & 0.0025 & 1.0000 & 0.0013 & 1.0000 & 1.0000 & 1.0000 & 0.0092 & - & - & - & - & - & - & - & - \\ 
Medical chatbot & 1.0000 & 1.0000 & 1.0000 & 1.0000 & 1.0000 & 1.0000 & 1.0000 & 1.0000 & - & - & - & - & - & - & - \\ 
Medical image analysis & 1.0000 & 1.0000 & 0.8862 & 1.0000 & 1.0000 & 0.4688 & 1.0000 & 1.0000 & 1.0000 & - & - & - & - & - & - \\ 
Medical text summarization & 1.0000 & 1.0000 & 1.0000 & 1.0000 & 0.1575 & 0.0013 & 1.0000 & 0.0002 & 1.0000 & 1.0000 & - & - & - & - & - \\ 
Natural language query interface & 1.0000 & 1.0000 & 1.0000 & 1.0000 & 1.0000 & 0.8862 & 1.0000 & 0.1172 & 1.0000 & 1.0000 & 1.0000 & - & - & - & - \\ 
Scheduling & $<$0.0001 & 0.0293 & 0.0002 & 0.2197 & 1.0000 & 1.0000 & 0.0002 & 1.0000 & 0.0586 & 0.2197 & $<$0.0001 & 0.0623 & - & - & - \\ 
Synthetic data generation & 0.1575 & 1.0000 & 0.0330 & 1.0000 & 1.0000 & 1.0000 & 0.0586 & 1.0000 & 1.0000 & 1.0000 & 0.0874 & 1.0000 & 1.0000 & - & - \\ 
Transcribing medical encounters & 1.0000 & 1.0000 & 1.0000 & 1.0000 & 1.0000 & 0.2820 & 1.0000 & 0.2820 & 1.0000 & 1.0000 & 1.0000 & 1.0000 & 0.0037 & 1.0000 & - \\ 
Other & 0.0333 & 1.0000 & 0.0188 & 1.0000 & 1.0000 & 1.0000 & 0.1027 & 1.0000 & 1.0000 & 1.0000 & 0.0655 & 1.0000 & 1.0000 & 1.0000 & 0.9045 \\ 
\bottomrule
\end{tabular}
}
\end{table}}

\newpage

{\small

\begin{table}[ht]
\centering
\caption{Mean ratings of important criteria when evaluating LLMs.}
\label{sup tab:evaluating mean-ratings}

\begin{tabular}{lr}
\toprule
\textbf{Use Case} & \textbf{Mean Rating} \\
\midrule
Use Case & 4.5278 \\
Accuracy & 4.4722 \\
Reproducible \& Consistent Answers & 4.4444 \\
Security \& Privacy Risk & 4.1944 \\
Legal \& Reputational Risk & 4.0556 \\
Cost & 3.9167 \\
Healthcare-Specific Models & 3.8611 \\
Explainability \& Transparency & 3.6111 \\
\bottomrule
\end{tabular}
\end{table}}

\newpage

{\small
\begin{table}[ht]
\centering
\caption{Mean ratings of significant limitation for current GenAI technology.}
\label{sup tab:genai mean-ratings}

\begin{tabular}{lr}
\toprule
\textbf{Use Case} & \textbf{Mean Rating} \\
\midrule
Compliance issue & 4.2222 \\
Lacks Accuracy & 4.1389 \\
Poses Legal, Security, or Reputational Risks & 4.0278 \\
Too Expensive & 3.9444 \\
Falls Short on Bias \& Fairness Requirements & 3.9167 \\
Not Built for Healthcare \& Life Science & 3.6389 \\
Not Tunable for Private Data or Use Cases & 3.5833 \\
\bottomrule
\end{tabular}
\end{table}}

\newpage

{\small
\begin{table}[ht]
\centering
\caption{Mean ratings of how much LLMs will impact each use case over the next 2-3 years.}
\label{sup tab:impact mean-ratings}

\begin{tabular}{lr}
\toprule
\textbf{Use Case} & \textbf{Mean Rating} \\
\midrule
Natural Language Query Interface & 4.5556 \\
Information Extraction / Data & 4.5000 \\
Medical Text Summarization & 4.4722 \\
Transcribing Medical Encounters & 4.3333 \\
Data Abstraction & 4.2778 \\
Medical Image Analysis & 4.1944 \\
Biomedical Research & 4.1111 \\
Machine Translation & 4.0278 \\
Clinical Coding / Chart Audit & 3.9167 \\
Medical Chatbot & 3.9167 \\
Answering Patient Questions & 3.7778 \\
De-identification & 3.6944 \\
Synthetic Data Generation & 3.5556 \\
Scheduling & 3.5278 \\
Drug Development & 3.3889 \\
Others & 3.2222 \\
\bottomrule
\end{tabular}
\end{table}}

\newpage

{\small
\begin{table}[!ht]
\caption{Results of post-hoc McNemar tests with Bonferroni correction for improvements observed since implementing GenAI solutions in the healthcare system.} \label{sup tab:LLM improvement}
\resizebox{\linewidth}{!}{
\begin{tabular}{r*{7}{>{\raggedleft\arraybackslash}p{5em}}}
\toprule
~ & Better patient engagement & Cost savings & Enhanced diagnostic accuracy & Faster decision-making processes & Improved patient outcomes & Increased operational efficiency & N/A - Have not started\\
\midrule
\rowcolor[gray]{.9}
\multicolumn{1}{l}{p-value} &&&&&&&\\
Cost savings & 0.1250 & - & - & - & - & - & -  \\
Enhanced diagnostic accuracy & 0.3750 & 0.6250 & - & - & - & - & - \\
Faster decision-making processes & 0.3750 & 0.0039 & 0.0414 & - & - & - & - \\
Improved patient outcomes & 0.0001 & 0.1338 & 0.3438 & 1.0000 & - & - & - \\
Increased operational efficiency & 0.0023 & 0.7905 & 0.0010 & 0.1892 & N/A & - & - \\
N/A - Have not started & 0.0391 & 1.0000 & 0.0414 & 0.2632 & N/A & N/A & - \\
Others & 0.6875 & 0.0034 & 0.3877 & N/A & N/A & N/A & N/A\\
\midrule
\rowcolor[gray]{.9}
\multicolumn{1}{l}{Corrrected p-value}  &&&&&&&\\
Cost savings & 1.0000 & - & - & - & - & - & - \\
Enhanced diagnostic accuracy & 1.0000 & 1.0000 & - & - & - & - & - \\
Faster decision-making processes & 1.0000 & 0.0820 & 0.8692 & - & - & - & - \\
Improved patient outcomes & 0.0026 & 1.0000 & 1.0000 & 1.0000 & - & - & - \\
Increased operational efficiency & 0.0493 & 1.0000 & 0.0205 & 1.0000 & N/A & - & - \\
N/A - Have not started & 0.8203 & 1.0000 & 0.8692 & 1.0000 & N/A & N/A & - \\
Others & 1.0000 & 0.0718 & 1.0000 & N/A & N/A & N/A & N/A\\
\bottomrule
\end{tabular}}
\end{table}}

\newpage

{\small
\begin{table}[!ht]
\caption{Results of post-hoc McNemar tests with Bonferroni correction for steps to test and improve LLM models. RLHF - Reinforcement Learning from Human Feedback.} \label{sup tab:LLM RLHF}
\resizebox{\linewidth}{!}{
\begin{tabular}{r*{10}{>{\raggedleft\arraybackslash}p{5em}}}
\toprule
~ & Adversarial testing & De-biasing tools and techniques & Guardrails & Human in the loop & Interpretability tools and techniques & Quantization and/or Pruning & Red Teaming & RLHF & Supervised fine-tuning & Others \\
\midrule
\rowcolor[gray]{.9}
\multicolumn{1}{l}{p-value} &&&&&&&&&&\\
De-biasing tools and techniques & 0.0074 & - & - & - & - & - & - & - & - & - \\ 
Guardrails & 0.0386 & 0.5811 & - & - & - & - & - & - & - & - \\ 
Human in the loop & $<$0.0001 & 0.0034 & 0.0001 & - & - & - & - & - & - & - \\ 
Interpretability tools and techniques & 0.0127 & 1.0000 & 0.6291 & 0.0034 & - & - & - & - & - & - \\ 
Quantization and/or Pruning & 1.0000 & 0.0034 & 0.0386 & $<$0.0001 & 0.0010 & - & - & - & - & - \\ 
Red Teaming & 0.6875 & 0.0010 & 0.0020 & $<$0.0001 & 0.0010 & 0.6875 & - & - & - & - \\ 
RLHF & 0.0654 & 0.3438 & 1.0000 & $<$0.0001 & 0.4240 & 0.0654 & 0.0225 & - & - & - \\ 
Supervised fine-tuning & 0.0005 & 1.0000 & 0.4545 & 0.0020 & 1.0000 & 0.0018 & 0.0005 & 0.2266 & - & - \\ 
Others & 0.7539 & 0.0044 & 0.0309 & $<$0.0001 & 0.0044 & 0.7266 & 1.0000 & 0.0352 & 0.0026 & - \\ 
None of the above & 0.1094 & 0.0002 & 0.0013 & $<$0.0001 & 0.0002 & 0.1094 & 0.2891 & 0.0023 & 0.0001 & 0.2891 \\ 
\midrule
\rowcolor[gray]{.9}
\multicolumn{1}{l}{Corrrected p-value} &&&&&&&&&&\\
De-biasing tools and techniques & 0.4062 & - & - & - & - & - & - & - & - & - \\ 
Guardrails & 1.0000 & 1.0000 & - & - & - & - & - & - & - & - \\ 
Human in the loop & $<$0.0001 & 0.1880 & 0.0067 & - & - & - & - & - & - & - \\ 
Interpretability tools and techniques & 0.6999 & 1.0000 & 1.0000 & 0.1880 & - & - & - & - & - & - \\ 
Quantization and/or Pruning & 1.0000 & 0.1880 & 1.0000 & $<$0.0001 & 0.0537 & - & - & - & - & - \\ 
Red Teaming & 1.0000 & 0.0537 & 0.1074 & $<$0.0001 & 0.0537 & 1.0000 & - & - & - & - \\ 
RLHF & 1.0000 & 1.0000 & 1.0000 & 0.0034 & 1.0000 & 1.0000 & 1.0000 & - & - & - \\ 
Supervised fine-tuning & 0.0269 & 1.0000 & 1.0000 & 0.1074 & 1.0000 & 0.1007 & 0.0285 & 1.0000 & - & - \\ 
Others & 1.0000 & 0.2434 & 1.0000 & 0.0005 & 0.2434 & 1.0000 & 1.0000 & 1.0000 & 0.1417 & - \\ 
None of the above & 1.0000 & 0.0122 & 0.0722 & $<$0.0001 & 0.0122 & 1.0000 & 1.0000 & 0.1292 & 0.0067 & 1.0000 \\
\bottomrule
\end{tabular}
}
\end{table}}

\newpage

{\small
\begin{table}[!ht]
\caption{Results of post-hoc McNemar tests with Bonferroni correction for type(s) of evaluations when deployed LLM solutions.} \label{sup tab:LLM evaluation}
\resizebox{\linewidth}{!}{
\begin{tabular}{r*{13}{>{\raggedleft\arraybackslash}p{5em}}}
\toprule
~ & Bias & Brand Voice & Explainability & Fairness & Freshness & Hallucinations / Disinformation & Ideological Leaning & Pruvate Data Leakage & Prompt Injection & Robustness & Sycophancy & Toxicity & Others \\
\midrule
\rowcolor[gray]{.9}
\multicolumn{1}{l}{p-value} &&&&&&&&&&&&&\\
Brand Voice & 0.7266 & - & - & - & - & - & - & - & - & - & - & - & - \\ 
Explainability & 0.0313 & 0.0313 & - & - & - & - & - & - & - & - & - & - & - \\ 
Fairness & 0.0034 & 0.0117 & 0.0625 & - & - & - & - & - & - & - & - & - & - \\ 
Freshness & 0.3438 & 0.3438 & 0.1094 & 0.2266 & - & - & - & - & - & - & - & - & - \\ 
Hallucinations / Disinformation & 0.0034 & 0.0117 & 0.2188 & 0.3438 & 1.0000 & - & - & - & - & - & - & - & - \\ 
Ideological Leaning & 0.0703 & 0.7744 & 1.0000 & $<$0.0001 & 0.0034 & 1.0000 & - & - & - & - & - & - & - \\ 
Private Data Leakage & 0.0010 & 0.0078 & 1.0000 & 0.0129 & 1.0000 & 0.0034 & 0.3438 & - & - & - & - & - & - \\ 
Prompt Injection & 1.0000 & 0.4545 & $<$0.0001 & $<$0.0001 & 0.1797 & 1.0000 & 0.5488 & N/A & - & - & - & - & -\\ 
Robustness & 0.0063 & 0.3593 & 1.0000 & 0.3438 & 0.3438 & 0.1797 & 1.0000 & N/A & N/A & - & - & - & - \\ 
Sycophancy & 0.2632 & 0.1797 & 0.1250 & 0.0001 & 0.1250 & 0.3438 & N/A & N/A & N/A & N/A & - & - & -\\ 
Toxicity & 0.1892 & 0.0020 & 1.0000 & 0.0525 & 0.1460 & 0.0063 & N/A & N/A & N/A & N/A & N/A & - & -\\ 
Others & 0.2188 & 0.1797 & 0.0010 & 0.0433 & 0.2891 & 0.2863 & N/A & N/A & N/A & N/A & N/A & N/A & -\\ 
None of the above & 0.0117 & 1.0000 & 1.0000 & 0.1250 & 1.0000 & 0.1892 & N/A & N/A & N/A & N/A & N/A & N/A & N/A\\
\midrule
\rowcolor[gray]{.9}
\multicolumn{1}{l}{Corrrected p-value} &&&&&&&&&&&&&\\
Brand Voice & 1.0000 & - & - & - & - & - & - & - & - & - & - & - & -  \\ 
Explainability & 1.0000 & 1.0000 & - & - & - & - & - & - & - & - & - & - & - \\ 
Fairness & 0.2256 & 0.7734 & 1.0000 & - & - & - & - & - & - & - & - & - & - \\ 
Freshness & 1.0000 & 1.0000 & 1.0000 & 1.0000 & - & - & - & - & - & - & - & - & - \\ 
Hallucinations / Disinformation & 0.2256 & 0.7734 & 1.0000 & 1.0000 & 1.0000 & - & - & - & - & - & - & - & - \\ 
Ideological Leaning & 1.0000 & 1.0000 & 1.0000 & 0.0040 & 0.2256 & 1.0000 & - & - & - & - & - & - & - \\ 
Private Data Leakage & 0.0645 & 0.5156 & 1.0000 & 0.8540 & 1.0000 & 0.2256 & 1.0000 & - & - & - & - & - & -\\ 
Prompt Injection & 1.0000 & 1.0000 & 0.0040 & 0.0040 & 1.0000 & 1.0000 & 1.0000 & N/A & - & - & - & - & -\\ 
Robustness & 0.4189 & 1.0000 & 1.0000 & 1.0000 & 1.0000 & 1.0000 & 1.0000 & N/A & N/A & - & - & - & -\\ 
Sycophancy & 1.0000 & 1.0000 & 1.0000 & 0.0081 & 1.0000 & 1.0000 & N/A & N/A & N/A & N/A & - & - & - \\ 
Toxicity & 1.0000 & 0.1289 & 1.0000 & 1.0000 & 1.0000 & 0.4189 & N/A & N/A & N/A & N/A & N/A & - & -\\ 
Others & 1.0000 & 1.0000 & 0.0645 & 1.0000 & 1.0000 & 1.0000 & N/A & N/A & N/A & N/A & N/A & N/A & - \\ 
None of the above & 0.7734 & 1.0000 & 1.0000 & 1.0000 & 1.0000 & 1.0000 & N/A & N/A & N/A & N/A & N/A & N/A & N/A \\
\bottomrule
\end{tabular}
}
\end{table}}

\newpage

{\small
\begin{table}[!ht]
\caption{Results of post-hoc McNemar tests with Bonferroni correction for challenges faced in integrating GenAI with existing systems.} \label{sup tab:LLM challenge}
\resizebox{\linewidth}{!}{
\begin{tabular}{r*{10}{>{\raggedleft\arraybackslash}p{5em}}}
\toprule
~ & Technical architecture and deployment & Data lifecycle management & Customization and integration features & Security & Scalability and performance & LLM output compliance & Legal and regulatory compliance & Monitoring and reporting & Financial and operational considerations & AI workforce development \\
\midrule
\rowcolor[gray]{.9}
\multicolumn{1}{l}{p-value} &&&&&&&&&&\\
Data lifecycle management & 0.1185 & - & - & - & - & - & - & - & - & - \\ 
Customization and integration features & 0.1796 & 1.0000 & - & - & - & - & - & - & - & - \\ 
Security & 0.1460 & 1.0000 & 1.0000 & - & - & - & - & - & - & - \\ 
Scalability and performance & 0.0005 & 0.3018 & 0.2101 & 0.2101 & - & - & - & - & - & - \\ 
LLM output compliance & 0.0005 & 0.3018 & 0.1796 & 0.1796 & 1.0000 & - & - & - & - & - \\ 
Legal and regulatory compliance & 0.0192 & 0.3438 & 0.2266 & 0.2668 & 1.0000 & 1.0000 & - & - & - & - \\ 
Monitoring and reporting & 0.0225 & 0.7744 & 0.5488 & 0.6072 & 0.5488 & 0.5488 & 0.7539 & - & - & - \\ 
Financial and operational considerations & 0.0117 & 0.7744 & 0.5488 & 0.5488 & 0.5488 & 0.6072 & 0.7539 & 1.0000 & - & - \\ 
AI workforce development & 0.0768 & 1.0000 & 0.7744 & 0.7905 & 0.3877 & 0.4545 & 0.5488 & 1.0000 & 1.0000 & - \\ 
Others & $<$0.0001 & 0.0004 & 0.0002 & 0.0002 & 0.0127 & 0.0127 & 0.0075 & 0.0026 & 0.0026 & 0.0015 \\
\midrule
\rowcolor[gray]{.9}
\multicolumn{1}{l}{Corrrected p-value} &&&&&&&&&&\\
Data lifecycle management & 1.0000 & - & - & - & - & - & - & - & - & - \\ 
Customization and integration features & 1.0000 & 1.0000 & - & - & - & - & - & - & - & - \\ 
Security & 1.0000 & 1.0000 & 1.0000 & - & - & - & - & - & - & - \\ 
Scalability and performance & 0.0269 & 1.0000 & 1.0000 & 1.0000 & - & - & - & - & - & - \\ 
LLM output compliance & 0.0269 & 1.0000 & 1.0000 & 1.0000 & 1.0000 & - & - & - & - & - \\ 
Legal and regulatory compliance & 1.0000 & 1.0000 & 1.0000 & 1.0000 & 1.0000 & 1.0000 & - & - & - & - \\ 
 Monitoring and reporting & 1.0000 & 1.0000 & 1.0000 & 1.0000 & 1.0000 & 1.0000 & 1.0000 & - & - & - \\ 
Financial and operational considerations & 0.6445 & 1.0000 & 1.0000 & 1.0000 & 1.0000 & 1.0000 & 1.0000 & 1.0000 & - & - \\ 
AI workforce development & 1.0000 & 1.0000 & 1.0000 & 1.0000 & 1.0000 & 1.0000 & 1.0000 & 1.0000 & 1.0000 & - \\ 
Others & 0.0008 & 0.0221 & 0.0122 & 0.0122 & 0.6999 & 0.6999 & 0.4146 & 0.1417 & 0.1417 & 0.0819 \\
\bottomrule
\end{tabular}
}
\end{table}}


\end{document}